\documentclass[11pt]{article} 
\usepackage{amsmath,amsthm, amssymb,amsfonts,amscd,mathabx}

\usepackage[colorlinks=true,linkcolor=blue,citecolor=red]{hyperref}
\usepackage[pdftex]{graphicx}                 

\graphicspath{{figs}{.}}

\usepackage{wrapfig}                               
\usepackage[font=footnotesize]{caption}                     
\usepackage{float}

\numberwithin{equation}{section}
\newtheorem{theorem}{Theorem}[section]
\newtheorem{lemma}[theorem]{Lemma}

\newtheorem{corollary}[theorem]{Corollary}
\newtheorem{hypothesis}[theorem]{Hypothesis}

\newenvironment{Proof}[1][.]%
{\begin{trivlist}\item[]\textbf{Proof#1 }}%
{\qed\end{trivlist}}

 {\begin{trivlist}\item[]\textbf{Acknowledgments }}{\end{trivlist}}

\theoremstyle{definition}

\newcommand{\R}{\mathbb{R}}
\newcommand{\C}{\mathbb{C}}

\newcommand{\Z}{\mathbb{Z}}

\newcommand{\Rmnum}[1]{\uppercase\expandafter{\romannumeral #1\relax}}

\newcommand{\rmO}{\mathrm{O}}

\newcommand{\rmd}{\mathrm{d}}
\newcommand{\rme}{\mathrm{e}}
\newcommand{\rmi}{\mathrm{i}}

\renewcommand{\Re}{\mathrm{Re}}
\renewcommand{\Im}{\mathrm{Im}}

\renewcommand{\leq}{\leqslant}
\renewcommand{\geq}{\geqslant}

\def\eps{\varepsilon}

\newcommand{\Tin}{{T_\mathrm{in}}}
\newcommand{\Tout}{{T_\mathrm{out}}}

\makeatletter\@addtoreset{figure}{section}\makeatother

\newfam\bifam
\font\tenbi=cmmib10 scaled \magstep1 \font\sevenbi=cmmib10 at 11pt
\font\fivebi=cmmib10 at 6pt \textfont\bifam = \tenbi
\scriptfont\bifam = \sevenbi \scriptscriptfont\bifam= \fivebi

\begin{document}

\begin{center}
{\fontsize{15}{15}\fontseries{b}\selectfont{
Slow passage through the Busse balloon -- predicting steps on the Eckhaus staircase 
}}\\[0.2in]
Anna Asch$^1$, Montie Avery$^2$, Anthony Cortez$^3$, and Arnd Scheel$^4$ \\[0.1in]
\textit{\footnotesize 
$^1$Department of Mathematics, Cornell University, Ithaca, NY, USA\\
$^2$Department of Mathematics and Statistics, Boston University, Boston, MA, USA\\
$^3$Department of Mathematics, Cal State University Fresno, Fresno, CA, USA\\
$^4$School of Mathematics, University of Minnesota, MN, USA 
}
\end{center}
\begin{abstract}
    Motivated by the impact of worsening climate conditions on vegetation patches, we study dynamic instabilities in an idealized Ginzburg-Landau model. Our main results predict time instances of sudden drops in wavenumber and the resulting target states. The changes in wavenumber correspond to the annihilation of individual vegetation patches when resources are scarce and cannot support the original number of patches. Drops happen well after the primary pattern has destabilized at the Eckhaus boundary and key to distinguishing between  the disappearance of 1,2, or more patches during the drop are complex spatio-temporal resonances in the linearization at the unstable pattern. We support our results with numerical simulations and expect  our results to be conceptually applicable universally near the Eckhaus boundary, in particular in more realistic models. 
\end{abstract}
\section{Introduction}

We study the evolution of spatially periodic patterns as system parameters vary slowly. Our motivation stems from ecological problems in which slowly varying parameters model worsening environmental conditions due to climate change. Of particular interest are dryland ecosystem models describing the interaction between vegetation density and available resources such as water. An important goal is to understand the process of desertification, predicting how much vegetation will remain as, for instance, average yearly rainfall decreases due to changing climate conditions. In ODE models which do not account for spatial variation of vegetation density or available resources, one often predicts \emph{tipping}, in which a sudden collapse from a vegetated to a desert state occurs once the yearly rainfall decreases past some critical value. One also often observes hysteresis in such models, so that it is difficult to reverse desertification once it has occurred even if average rainfall begins to recover~\cite{ArjenTipping}. 

There has been great recent interest~\cite{BastiaansenDoelman,BastiaansenEtAl, PaulInstabilities, MehronFronts,rietkerk_2004,meron_2019,ArjenTipping} in understanding the role of spatial dependence of the distribution of vegetation and resources on the ecosystem dynamics and eventual collapse. In particular, it was recently observed in that in spatially extended models, a Turing instability of the uniformly vegetated state may occur prior to reaching the tipping point in the spatially independent system; see~\cite{ArjenTipping} and references therein. Uniform vegetation then gives way to periodic vegetation patches, and the pattern now evolves through the space of stable periodic patterns, called the \emph{Busse balloon}; see~\cite{busse1,Busse_1978} and~\cite[Ch.IIIB2b]{cross_hohenberg}. In some cases, the system may avoid tipping completely, with stable periodic vegetation patterns persisting past the former tipping point. This ``Turing before tipping'' mechanism then increases the resilience of the ecosystem as it avoids total collapse and may in principle recover by evolving back through the Busse balloon of stable patterns if environmental conditions begin to improve.

\begin{figure}[h]
    \centering
    \includegraphics[width=0.8\textwidth]{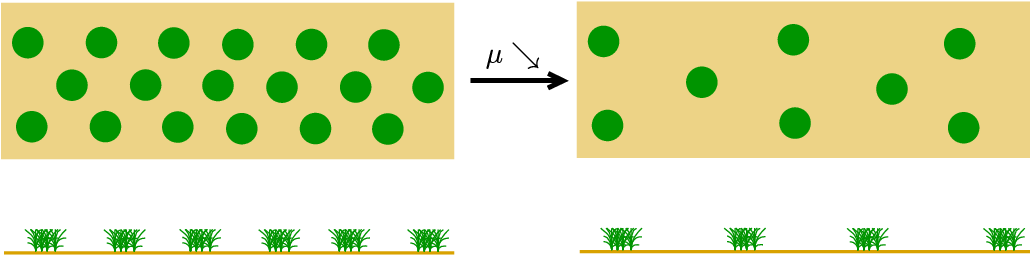}\\[0.3in]
    \includegraphics[width=0.8\textwidth]{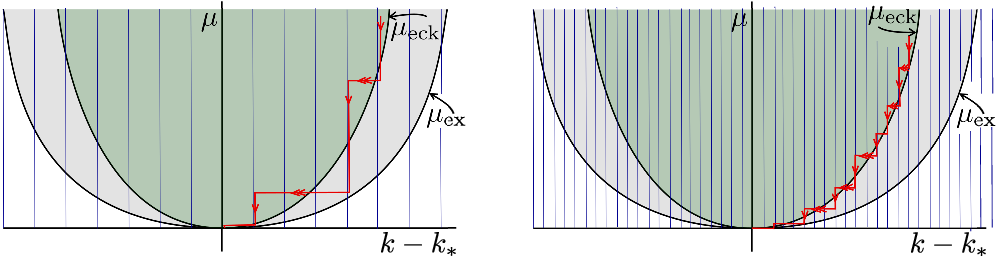}\\
    \caption{
Top: Schematic depiction of vegetation patches in one and two dimensional domains, and drops of density as parameters evolve slowly;
Bottom: Schematic depiction of the Busse balloon near a Turing instability with parameter $\mu$ vertical, wavenumber horizontal. Wavenumbers of equilibria are quantized due to imposed spatially periodic boundary conditions with  small (left) and large (right) period; see for example the discussion in~\eqref{e: GL conj}. Patterns exist above $\mu_\mathrm{ex}$ but are stable only above $\mu_\mathrm{eck}$. Red curves are schematic sample paths of observed wavenumbers when the parameter $\mu$ is slowly decreased, evolving in a staircase along the Eckhaus boundary. See also Figure~\ref{f:staircase} for numerical simulations.
}
    \label{f:busse}
\end{figure}

The purpose of this work is to better understand the evolution of patterns in and beyond the Busse balloon as system parameters vary slowly. One expects that the pattern will vary continuously, slightly adjusting the amplitude and shape of vegetation patches, until it reaches the boundary of the Busse balloon, at which point the pattern becomes unstable in the  system with frozen parameters. In ecological terms, the available resources are no longer sufficient to support the present number of vegetation patches, so the system ``sacrifices'' some number of vegetation patches, giving way to a new stable pattern with fewer vegetation patches. This self-organized sacrifice intuitively requires coordination between the different vegetation patches, all of which experience the same scarcity of resources. In parameter space, this crisis is  represented by a sudden jump from the boundary to the interior of the Busse balloon, to a pattern with a lower wavenumber; see Fig.~\ref{f:busse} for an illustration. 

\begin{figure}
    \centering
    \includegraphics[width=0.4\textwidth]{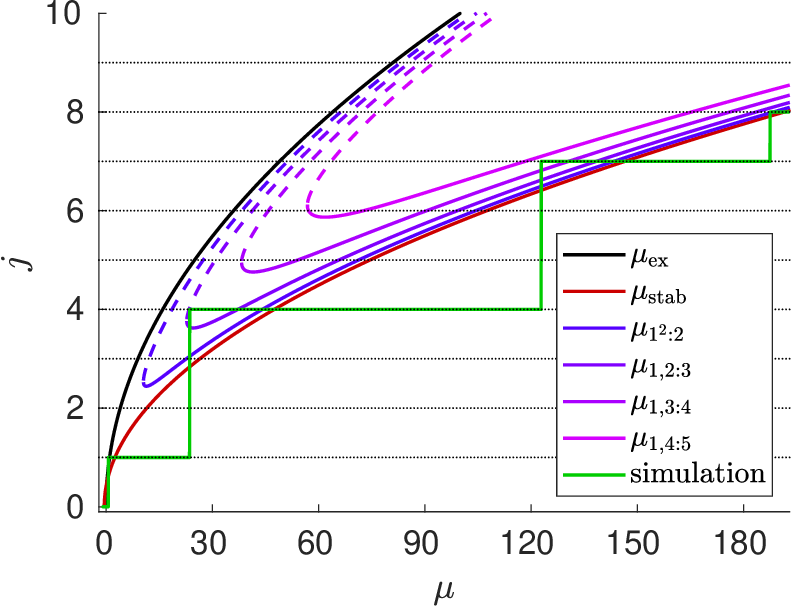}\qquad
    \includegraphics[width=0.4\textwidth]{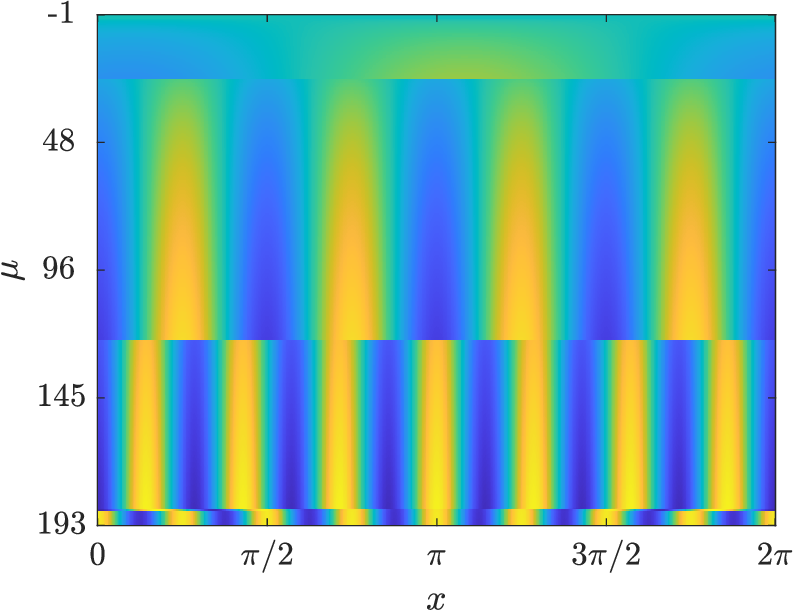}
    \caption{Left: Existence and stability boundaries, as well as spatio-temporal resonances responsible for cross-overs in the drop number changes, together with mode number of a sample trajectory. Resonance curves $\mu_{1^2:2}$, $\mu_{1,2:3}$, etc. correspond to parameter values where $\lambda_1+\lambda_1=\lambda_2$, $\lambda_1+\lambda_2=\lambda_3$, and so forth. In the region between solid and dashed part of the curves, $\lambda_{j+1}>\lambda_1+\lambda_j$; see \S\ref{ss:main} for details on resonances and their relevance.  Right: Space-time plot of $\Re(A)$ of the same sample trajectory, with time axis represented in terms of the slowly varying $\mu$, varying up until $\mu<0$ and no nontrivial equilibria exist.}
    \label{fig:sample staircase}
\end{figure}

While this picture of ``bouncing off the boundary of the Busse balloon'' is somewhat intuitive and is born out in numerical simulations (see for instance Fig.~\ref{fig:sample staircase}), there do not appear to be theoretical predictions of these dynamics of evolution through the Busse balloon in simple pattern-forming models\footnote{See however~\cite{BastiaansenDoelman} for predictions of the evolution in the presence of drift and heterogeneity.}. For instance, while it is intuitive that one must jump to another stable pattern once the original underlying pattern becomes unstable past the boundary of the Busse balloon, it is not clear how many vegetation patches will be lost in such a transition. A natural goal is then to answer the following question.
\begin{center}
    \emph{Given an initial periodic pattern with small fluctuations, can we predict the evolution of the number of patches as parameters vary slowly? In particular, can we predict the time at which the number of patches will change, and how much it will drop by?} 
\end{center}
We formulated our motivation in terms of dryland vegetation patches, but we believe that there are many settings in which this evolution of patterns with varying parameters is at the center of the description of dynamics; see for instance studies of Turing patterns in growing domains~\cite{MR4350741} or the analysis of melting-boundary convection in~\cite{MR2850970}.

\begin{figure}
    \centering
    \includegraphics[width=0.6\textwidth]{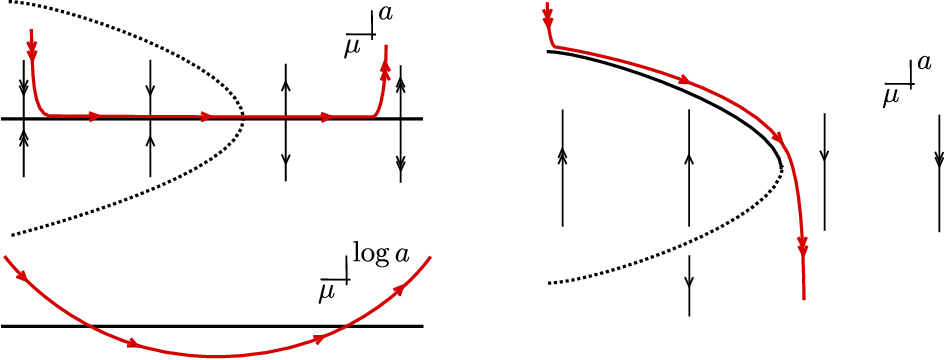}
    \caption{Delay of bifurcation in pitchfork (left) and saddle-node bifurcation when the parameter is slowly varied. Left, top: Slow passage through a subcritical pitchfork bifurcation, $a'=\mu a + a^3,\ \mu'=\eps$, yields $\rmO(1)$ delay in parameter space of the departure from the unstable state. Left, bottom: The reason for the delay in the pitchfork bifurcation is an accumulated exponential smallness from the dynamics in the stable regime, here shown in a schematic log plot of the amplitude. Right: In a slow passage near a fold, $a'=-\mu-a^2,\ \mu'=\eps$, the delay is small, $\rmO(\eps^{1/3})$ in the parameter. See text for details on how our results relate to the delay in the pitchfork bifurcation. }
    \label{f:pf_concept}
\end{figure}

The question embeds naturally into the larger set of problems known as \emph{delay of instability}; see for instance ~\cite{MR1166998,HKSW_2016,MR2552125,MR1846577}, in particular the early work in~\cite{neishtadt1985asymptotic}, and see~\cite{MR4132535,crampin1999reaction,PhysRevE.94.022219,goh2023growing} for dynamic bifurcations in the context of pattern-forming instabilities. Such delays manifest themselves in contexts where a branch of equilibria undergoes a bifurcation and destabilizes, or vanishes entirely. ``Dynamically'' passing through this bifurcation diagram by slowly changing the parameter with rate $\eps$, one expects to follow the ``static'' picture, that is, one expects to track the stable branch and, in the absence of the stable branch, a strong unstable manifold of the equilibrium at criticality; see Fig.~\ref{f:pf_concept}. In the simplest case of a saddle-node bifurcation, the solution stays near the remnant of the equilibria for a time $\eps^{-1/3}$~\cite{ks1}. Similar results are available in the case of pitchfork and transcritical bifurcations when the parameter change includes a generic disturbance of the trivial branch~\cite{ks2}. If the trivial branch is undisturbed, the delay is of order $\eps^{-1}$, that is, of order one in the parameter, with take-off point determined by the initial parameter value. Of course such a scenario is global in the sense that it requires assumptions on the system far away from the bifurcation point, at which point also other eigenvalues might be relevant; see for instance~\cite{Kaklamanos2023} for an analysis of two such eigenvalues interacting and~\cite{hummel2022geometric} for a case with continuous spectrum crossing. More robust delays occur in Hopf bifurcations~\cite{nei1,nei2,HKSW_2016}, where recent work, closer in spirit to our work here, also explores a PDE setting where many modes destabilize instantaneously~\cite{gkv_2022,kv_2018}.

Our setting relates to the global aspect of the delay of instability in pitchfork bifurcations. In fact, the Eckhaus instability, which generically describes the boundary of the Busse balloon near a Turing bifurcation, manifests itself in bounded domains as a subcritical pitchfork bifurcation. With this analogy,  we predict delays of order one in the parameter before leaving a fixed pattern. A simple linear calculation indeed  correctly  predicts the take-off point through an integral criterion: take-off points are found by requiring that the average of the marginally stable eigenvalue along part of the branch followed by the solution is zero. Intuitively, the cumulative exponential decay in the stable regime is ``spent'' in the unstable regime where the solution grows exponentially, until both cancel and the solution reaches finite distance from the equilibrium branch  again.
Beyond this simple linear calculation that we corroborate and supplement with global information, our key insight here is that the global nature of the delay makes it necessary to take into account other potentially unstable modes. Our results in this direction answer the main question posed above through \emph{linear predictions based on the presence of nonlinear spatio-temporal resonances}. We predict, depending on the initial parameter value, the time when the solution leaves a neighborhood of the equilibrium and the new equilibrium that the solution approaches. Since in fact the number of maxima (or vegetation patches, in this particular scenario) drops, we refer to this critical time as the \emph{drop time}, to the associated parameter value as the \emph{drop parameter value}, and to the decrease in the number of maxima as the \emph{drop number}. Our results predict drop numbers in a range from 1 to 4 and associated drop times. The predictions are based on spatio-temporal resonances in the complex plane and are reminiscent of criteria for front invasion speeds and transitions from convective to absolute instability in large bounded domains as explored in~\cite{fhs_2017,adss_2021} and point to a potentially more comprehensive theory of spatio-temporal instabilities in extended systems. We emphasize that the results here address one particular, albeit ubiquitous example of instability, the long-wavelength Eckhaus destabilization. Boundaries of the Busse balloon in systems far from equilibrium often involve different instability modes, in particular period-doublings; see~\cite{MR3817752,MR2836552} for examples and see~\cite{MR2355524} for a conceptual view on instabilities of Turing patterns in one-dimensional systems.

To be specific, we study this question here in the Ginzburg-Landau equation
\begin{align}
A_t = A_{xx} + \mu(\eps t) A - A |A|^2, \quad A = A(x,t) \in \C, \quad x\in \R, \quad t > 0. \label{e: GL}
\end{align}
The Ginzburg-Landau equation is one of the simplest models of pattern-forming systems, and a universal amplitude equation which describes spatiotemporal dynamics near a Turing instability~\cite{cross_hohenberg,newell_whitehead_1969}. It is therefore a natural starting point to understand the evolution of periodic patterns through the Busse balloon. In the Ginzburg-Landau equation, the boundary of the Busse balloon is determined by the \emph{Eckhaus instability}; see~\cite[iVA1a(ii)]{cross_hohenberg} for background,~\cite{KRAMER1988212,PhysRevLett.81.18} for a study of the dynamics of the instability,~\cite{eckhaus_fluctuations,Tarlie_1998} for the effect of noise,~\cite{zimmermann_2021} for finite-size effects, 
and  \S\ref{s: Eckhaus static} below for a basic review. We allow the linear coefficient $\mu$ to vary slowly in time, with time scale $\eps$, and consider the problem on bounded domains with periodic boundary conditions. We will often make comparisons to the \emph{static problem}, where $\mu$ does not vary in time. We emphasize here that when interpreting results obtained for the Ginzburg-Landau equation in the context of a Turing instability, one needs to account for the fact $A(x,t)$ is an amplitude-phase modulation, so that actual solutions are of the form $A(\eps x,\eps^2 t)\rme^{\rmi x}$, say. In particular, the preferred \emph{constant} state $A\equiv const$ in Ginzburg-Landau corresponds to a preferred patterned state in the original model.

The equation~\eqref{e: GL} with $\eps=0$, $\mu(\eps t)=\mu>0$ fixed,  possesses equilibria $A_{*}(x;j)=\sqrt{\mu-j^2}\rme^{\rmi jx}$ for all $|j|\leq \sqrt{\mu}$. We are interested in the fate of these ``patterned'' equilibria for a fixed $j$ when the parameter $\mu$ is slowly decreased and, as it turns out, the equilibrium is destabilized. We wish to consider instabilities on a fixed-size domain which we assume, after possibly rescaling $x,t$, and $|A|$,  to be $x\in[0,2\pi]$. To accommodate an initially fixed equilibrium with arbitrary $j$ we impose $2\pi$-twist-periodic boundary conditions, relative to this equilibrium,
\begin{align}
    A(2 \pi, t) = \rme^{2\pi  \rmi j} A(0, t), \quad A_x(2 \pi, t) = \rme^{2 \pi  \rmi j} A_x (0, t). \label{e: A bc}
\end{align}
This allows us to treat $j$ as a continuous  rather than a quantized parameter.

Next, defining a new variable $B$ through $A(x,t) = \rme^{\rmi j x} B(x,t)$, we find that $B$ solves the conjugated equation
\begin{align}
    B_t = (\partial_x + \rmi j)^2 B + \mu(\eps t) B - B |B|^2, \label{e: GL conj}
\end{align}
with periodic boundary conditions
\begin{align}
    B(2 \pi, t) = B(0, t), \quad B_x (2 \pi, t) = B_x (0, t). \label{e:bcp}
\end{align}
This equation has a family of constant solutions
\begin{equation}\label{e:Ej}
E_j=\{\rme^{\rmi\varphi}B_*(j),\varphi\in[0,2\pi)\},\quad B_*(j) =\sqrt{\mu - j^2},
\end{equation}
modulated solutions
\begin{equation}\label{e:Ej'}
E_{j'}=\{\rme^{\rmi\varphi}B_*(j') \rme^{i(j'-j)x},\varphi\in[0,2\pi)\},\quad B_*(j') =\sqrt{\mu - j'^2},
\end{equation}
and the trivial solution $B_*=0$. Drop numbers as described above, when, say, $j>0$,   are now simply the differences $j-j'$ for a trajectory that hovers subsequently first near an equilibrium $E_j$ and later near an equilibrium $E_{j'}$ with $j>|j'|$. Together with the drop number, we also wish to predict the drop time $t_*$, when a solution leaves a small fixed neighborhood of $E_{j}$, or, more naturally, the associated drop parameter value $\mu(\eps t_*)$. The main insight of the analysis presented in this paper can be roughly described is as follows.


\noindent \textbf{Main results.} \emph{Consider solutions to~\eqref{e: GL conj} with periodic boundary conditions ~\eqref{e:bcp}, and with initial conditions close to $E_j$ for some $j$, and assume that $E_j$ is initially stable for $\mu = \mu_0$. Under conceptual assumptions on the existence of connecting orbits in the static problem, drop times and drop numbers as $\mu$ varies are determined by resonance criteria for eigenvalues, and associated integral conditions summarized in \S\ref{ss:main}. }

\begin{figure}
    \centering
    \includegraphics[width=\textwidth]{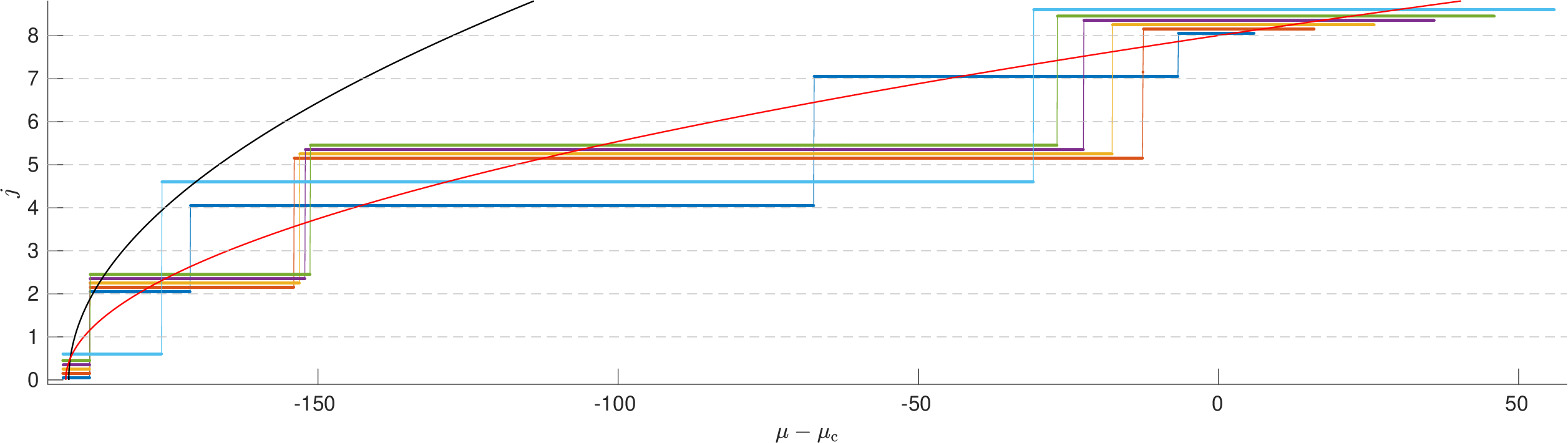}
    \caption{Simulations of~\eqref{e: GL} with initial condition $A_*(x;j)$, $j=8$ and $\eps=0.05$, with initial $\mu$-values $\mu_\mathrm{in}=\mu_\mathrm{c}+\hat{\mu}$, $\hat{\mu}=6,16,\ldots,56$. Plotted are winding numbers of $A$ as time progresses and $\mu$ decreases, over the parameter value $\mu$ at a given time instance. Trajectories crisscross the Eckhaus boundary in a staircase pattern that depends on the initial parameter value.  The value $\hat{\mu}=0$ corresponds to the onset of the Eckhaus instability at $j=8$. Initial conditions further away from the instability lead to later drops and higher drop numbers. Fractions of 1 are added to show all itineraries simultaneously, so that the actual current winding number is the largest integer below the plotted curve. Also shown are the boundary of instability (red) and the boundary of existence of equilibria (black) with a given $j$; see \S\ref{s: Eckhaus static} for details. Note that the figure is reflected along the diagonal when compared to Fig.~\ref{f:busse}, using the traditional bifurcation-theoretic depiction of phase-space versus parameter also used in Fig.~\ref{f:pf_concept} .  }
    \label{f:staircase}
\end{figure}

As a consequence, our results predict the drops in the intricate web of winding numbers $j$ observed over time, when only the initial parameter value is varied; see Fig.~\ref{f:staircase} for this increase of the drop with distance from criticality and for the ensuing sequence of drop events. 

\noindent \textbf{Acknowledgements.} The material here is based on work supported by the National Science
Foundation. All authors were supported by NSF-DMS-1907391, while M.A. was additionally supported by NSF-GRFP-0074041 and NSF-DMS-2202714. Any opinions, findings, and conclusions or recommendations expressed in this material are those of the authors and do not
necessarily reflect the views of the National Science Foundation.


\section{The Eckhaus instability in the static problem}\label{s: Eckhaus static}

In this section, we recall stability of periodic patterns in the static problem
\begin{align}
    B_t = (\partial_x + \rmi j)^2 B + \mu B - B |B|^2, \label{e: GL conj static}
\end{align}
with $\mu$ fixed and with $2\pi$-periodic boundary conditions.  Dynamics of~\eqref{e: GL conj static} are  influenced by its symmetries, which we now list for reference:
\begin{itemize}
    \item (spatial translation) $T_{x_0}: B(x,t) \mapsto B(x+x_0, t)$, viewing $x$ on the circle $\R / 2 \pi \Z$;
    \item (reflection) $T_-: B(x,t) \mapsto \rme^{-2\rmi jx} B(-x,t)$;
    \item (gauge rotation) $R_{\phi}: B(x,t) \mapsto \rme^{i\phi} B(x,t)$ for $\phi \in \R/2\pi\Z$;
    \item (complex reflection) $S_-:B(x,t)\mapsto \rme^{-2\rmi jx}\bar{B}(x,t)$.
\end{itemize}
From the discussion above, we recall that we are interested in the stability properties of the spatially constant equilibria $E_{j}$, defined in~\eqref{e:Ej}.
We remark that one commonly studies stability properties on the unbounded real line rather than on a fixed, possibly large finite domain.  Separating into phase and amplitude dynamics and linearizing, one there finds that the $E_{j}$ are linearly stable provided $\mu > 3 j^2$, and the long wavelength instability at $\mu = 3 j^2$ is referred to as the \emph{Eckhaus instability}; see for instance~\cite[Chapter 8.2]{hoyle} for a review. Our focus on finite domains here gives well known corrections to this threshold. The restriction to finite domains here appears essential to the techniques used below.

\subsection{Local stability analysis} 
\begin{figure}
    \centering
    \includegraphics[width=0.4\textwidth]{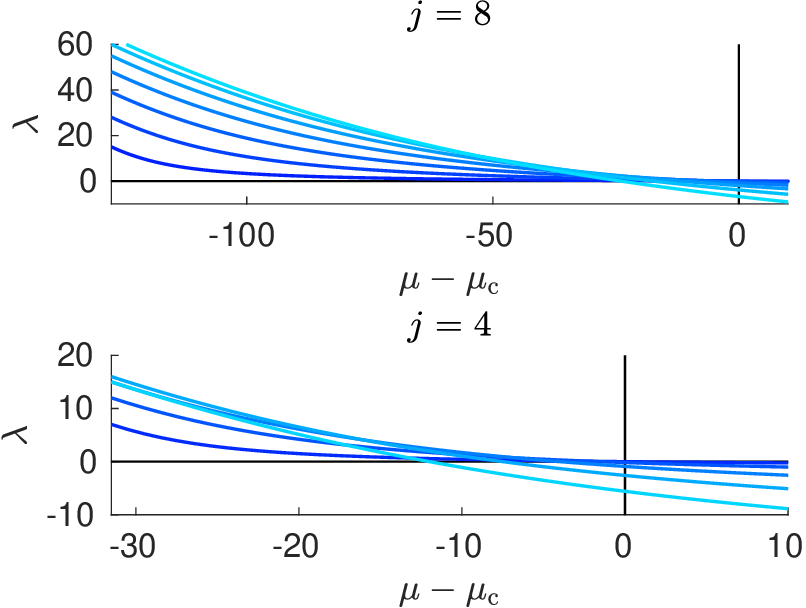}\qquad
    \includegraphics[width=0.4\textwidth]{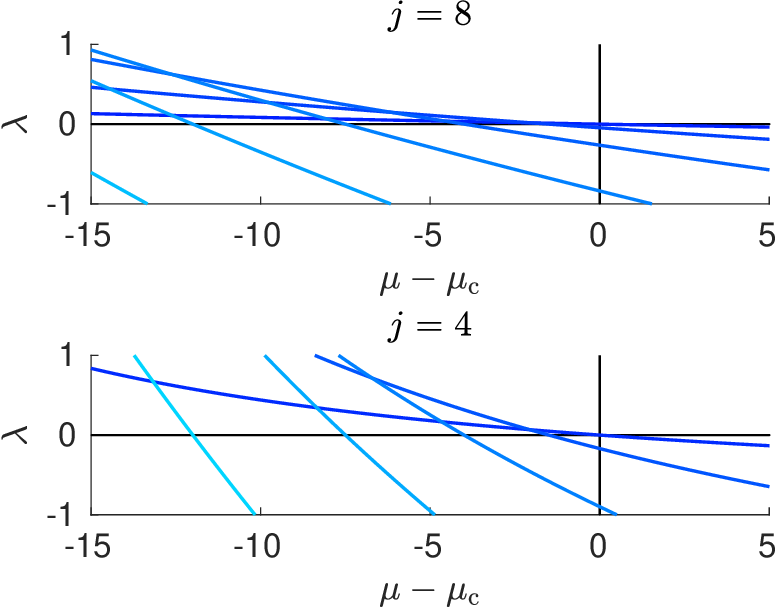}
    \caption{Eigenvalues of the linearization at $E_j$ with $j=8$ (top) and $j=4$ (bottom), for parameter values $\mu=j^2$ at onset to $\mu_\mathrm{c}+10$. Enlargement near criticality (right) reveals the intricate crossovers, that are explicit in the limit of large $j$. Eigenvalues are shown from $\ell=1$ in dark blue to $\ell=7$ (top) and  $\ell=5$ (bottom), respectively, in light blue.}
    \label{fig:evs}
\end{figure}

The Eckhaus instability in finite domains was analyzed in~\cite{BarkleyTuckerman}, and we briefly review the basic linear analysis here. Separating~\eqref{e: GL conj static} into real and imaginary parts $B = u + iv$ and linearizing about $(u,v) = (B_*(j), 0)$, we find
\begin{align*}
    u_t &= u_{xx} - 2 j v_x - 2 (\mu - j^2) u \\
    v_t &= v_{xx} + 2 j u_x. 
\end{align*}
The linearization is block diagonal in Fourier modes
 $   u(x) = \sum_{\ell \in \Z} u_\ell \rme^{\rmi \ell x},\ v(x) = \sum_{\ell \in \Z} v_\ell \rme^{\rmi \ell x}
$,
\begin{align}
    \begin{pmatrix}
    \dot{u_\ell} \\
    \dot{v_\ell} 
    \end{pmatrix}
    = \begin{pmatrix}
    -\ell^2 - 2( \mu -j^2) & - 2 \rmi j \ell \\
    2  \rmi j \ell & -\ell^2 
    \end{pmatrix} \begin{pmatrix} u_\ell \\ v_\ell \end{pmatrix}.
\end{align}
This matrix is self-adjoint, a reminder that the Ginzburg-Landau equation is a gradient flow. It has a negative trace, so that it always has at least one negative real eigenvalue. From a short calculation, one finds that the other eigenvalue is positive when
\begin{align}
    \mu < \mu_{\ell,\mathrm{c}} := 3j^2 - \frac{\ell^2}{2} \label{e: mu crit}
\end{align}
for $\ell \neq 0$, while the $\ell = 0$ block always has one zero and one negative eigenvalue. As $\mu$ decreases, the instability is therefore always first seen in the $\ell = \pm 1$ mode. The eigenvalue which becomes unstable past $\mu_{\ell,\mathrm{c}}$ is given by
\begin{align}
 \lambda_{\ell,+}(\mu) = - \ell^2 - (\mu-j^2) + \sqrt{4\ell^2 j^2 + (\mu-j^2)^2}.
\end{align}
For notational simplicity, we usually suppress the dependence of $\lambda_{\ell,+}(\mu)$ and $\mu_{\ell,\mathrm{c}}$ on the base wavenumber $j$. To further simplify notation, we write $\lambda_\ell$ instead of $\lambda_{\ell,+}$ since the eigenvalue $\lambda_{\ell,-}$ will be irrelevant for the discussion in this work. We also write simply $\mu_\mathrm{c}:=\mu_\mathrm{1,c}$ for the boundary of stability, noting that modes with $|\ell| \neq 1$ destabilize for smaller values of $\mu$, reflecting the side-band nature of the Eckhaus instability.
Eigenvectors are 
\[
(u_{\ell},v_{\ell})(\mu)=\left(j^2-\mu+\sqrt{4\ell^2 j+(\mu-j^2)^2},2\rmi\ell j\right). 
\]
Note that, since the equilibrium is fixed under translations $T_{x_0}$ and the reflection $T_-S_-: B(x)\mapsto\bar{B}(-x)$, the linearization commutes with these symmetries which henceforth act on eigenspaces (and on the flows on invariant manifolds tangent to those eigenspaces). 
Since the algebraic expression for eigenvalues and eigenvectors are somewhat cumbersome, we will later verify assumptions explicitly in the limit of large $j$, where we set $\mu=\mu_\mathrm{c}+\hat{\mu}$, and find
\begin{align}
    \lambda_{\ell}(\mu)=\frac{-\ell^2}{2}\left(\frac{\ell^2}{2}-\frac{1}{2}+\hat{\mu}\right)j^{-2}+\rmO\left(j^{-4}\right),\qquad 
    (u_{\ell},v_{\ell})(\mu)=\left(\frac{\ell}{2}+\rmO\left(j^{-2}\right),\rmi j\right),\label{e:lamlj}
\end{align} 
also writing   $e_{\ell}(\mu)$ short for the normalized eigenvector; see Fig.~\ref{fig:evs} for an illustration of the dependence of eigenvalues on $\mu$.
The stability threshold $\hat{\mu}=0$ corresponds to the Eckhaus boundary $\mu = 3 j^2$ in the infinite domain, but with the finite size correction $-\frac{1}{2}$
 Note that since $u$ and $v$ are real, we have $u_\ell = \overline{u_{-\ell}}, v_\ell = \overline{v_{-\ell}}$, and in particular $\lambda_{\ell}(\mu) = \lambda_{-\ell}(\mu).$

\subsection{Connecting orbits} 
To understand the evolution of periodic patterns in the dynamic problem, we are interested not only in stability boundaries in the static problem, but also in \emph{connecting orbits}. That is, we ask the following question: if we start with an initial condition $B_0$ near the equilibrium $E_j$ which is Eckhaus unstable, where does the solution $B(t)$ go as $t \to \infty$?

The complex Ginzburg-Landau equation is in fact a gradient flow in $L^2([0,2\pi),\C)$,
\begin{equation}\label{e:cglenergy}
 B_t=-\nabla_{L^2}\mathcal{E}[B],\qquad \text{ with energy }\quad     \mathcal{E}[B] = \frac{1}{2 \pi}\int_0^{2 \pi} \left(\frac{1}{2} |B_x+\rmi jB|^2 - \frac{\mu}{2} |B|^2 + \frac{1}{4} |B|^4 \right)\, \rmd x. 
\end{equation}
As a consequence, the solution $B(t)$ necessarily converges to an equilibrium with energy lower than $E_j$. A direct calculation shows that from this perspective, out of all equilibria $E_k$, only wavenumbers  $|k| < j$ are eligible, and one may suspect that solutions $B(t)$ originating near $E_j$ converge to a specific $E_k$. Unfortunately, it seems difficult to establish analytically which equilibrium $E_k$ is selected in this sense. In fact, there are also spatially patterned equilibria other than the $E_k$, believed to all be unstable as they limit, in a large-domain limit, onto unstable phase defect solutions such as
\begin{equation}\label{e: defect}
A_\mathrm{d}(x;k)=\left(\sqrt{2}k+\rmi \sqrt{1-3k^2}\tanh(\sqrt{1-3k^2}x/\sqrt{2})\right)\rme^{\rmi k x};
\end{equation}
see~\cite{MR3376770}.  Some, for our purposes rather restrictive, results have been obtained in~\cite{EckmannGallay}, where heteroclinic orbits between two distinct equilibrium sets $E_j$ and $E_{j-k}$ have been constructed. The arguments there are based on the the variational structure~\eqref{e:cglenergy}, roughly demonstrating that there is in fact a unique candidate for connecting orbits with given co-periodicity. In our contexts, the results only give conclusive information in the case  $\frac{1}{2} < j < 1$, in which case a generic perturbation of $E_j$ will converge to $E_{j-1}$.

On the other hand, the fate of perturbations is rather straightforward to establish numerically. We distill our findings in the following conceptual  assumption about the static problem, which we will later rely on to draw conclusions in the problem with slowly varying parameters. 


First note that for a generic $\mu$, the eigenvalues $\lambda_{\ell} (\mu)$ will be distinct up to the symmetry $\lambda_{\ell} (\mu) = \lambda_{-\ell} (\mu)$. The special crossover points at which this condition is violated, that is, at which eigenvalues for different wavenumbers $\ell$ collide, will play a role later but are excluded in the following discussion. We denote the crossover point, at which $\lambda_{\ell}(\mu) = \lambda_{k}(\mu)$, by $\mu = \mu_{\ell: k}$. 
Away from these crossover points, we let $\rme^\mathrm{u}_\ell$ denote the real two-dimensional eigenspace associated with the unstable eigenvalues $\lambda_{\pm\ell}(\mu)$.

\begin{figure}
    \centering\includegraphics[width=0.65\textwidth]{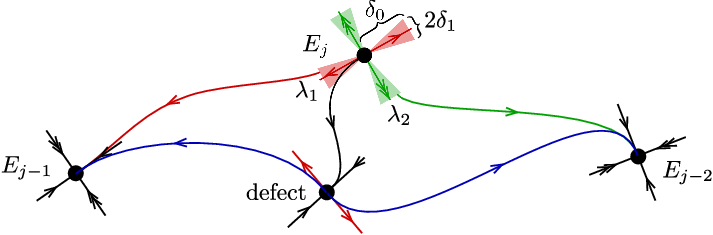}
    \caption{Schematic of the picture implied by Hypothesis~\ref{hyp: mode conservation}: The unstable manifold of $E_j$, here 2-dimensional, contains open sets of orbits that connect to $E_{j-1}$ and $E_{j-2}$, respectively. The hypothesis guarantees that the connecting orbits to $E_{j-1}$ and $E_{j-2}$, respectively, contain  the caps of conical regions around the respective eigenspaces. In particular, trajectories in the boundary between connecting orbits to $E_{j-1}$ and $E_{j-2}$ are not arbitrarily close to the eigenspaces. It seems plausible that the boundary of validity of the hypothesis is related to trajectories on a codimension-one stable manifold of saddle equilibria, here referred to as ``defect'', since equilibria of the form~\eqref{e: defect} are natural candidates for such a role in the setting of an unbounded domain. }\label{f:mc}
\end{figure}

The following hypothesis is sometimes colloquially referred to as node conservation or linear mode selection.
\begin{hypothesis}[Linear mode selection]\label{hyp: mode conservation}
For $0<\delta_1\ll\delta_0\ll 1$ sufficiently small fixed,  trajectories with initial conditions given as a perturbation of  $E_{j}$ of size $\delta_0$ will limit on $E_{j-\ell_*}$ if their amplitude is maximal in the unstable eigenspace with index $\ell_*$. To be precise,  writing the projection of the initial condition  on the unstable subspace of $E_j$ as a linear combination $\sum\alpha_\ell e_{\ell} \rme^{\rmi\ell x}$, we have $\delta_0=|\alpha_{\ell_*}|\gg{\delta_1}=\max_{\ell\neq\ell_*}|\alpha_\ell|$. Here, $e_{\ell}$ is the normalized eigenvector to the eigenvalue $\lambda_{\ell}$ from~\eqref{e:lamlj}; see Fig.~\ref{f:mc} for a schematic.
\end{hypothesis}

To explain the basic intuition, notice that perturbations of the constant equilibrium in the direction of the eigenvector $\lambda_{\ell}$, that is perturbations where $\delta_1=0$, have a sinusoidal shape and winding number $\ell$ relative to $E_j$. The hypothesis states that as the amplitude grows, this ``modal'' shape is preserved and terminates in the equilibrium $E_{j-\ell}$, with winding number $-\ell$ relative to the origin.

\begin{figure}
    \centering
    \includegraphics[width=0.4\textwidth]{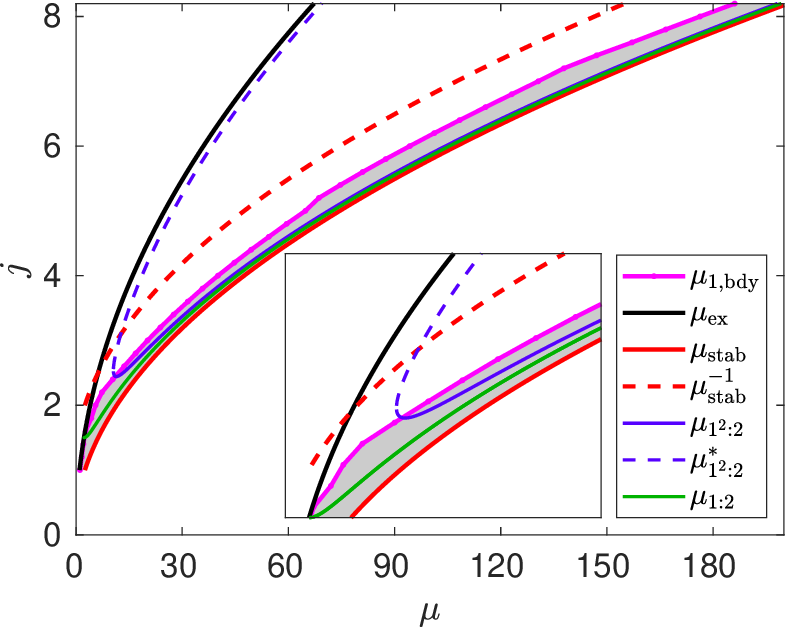} \qquad
    \includegraphics[width=0.4\textwidth]{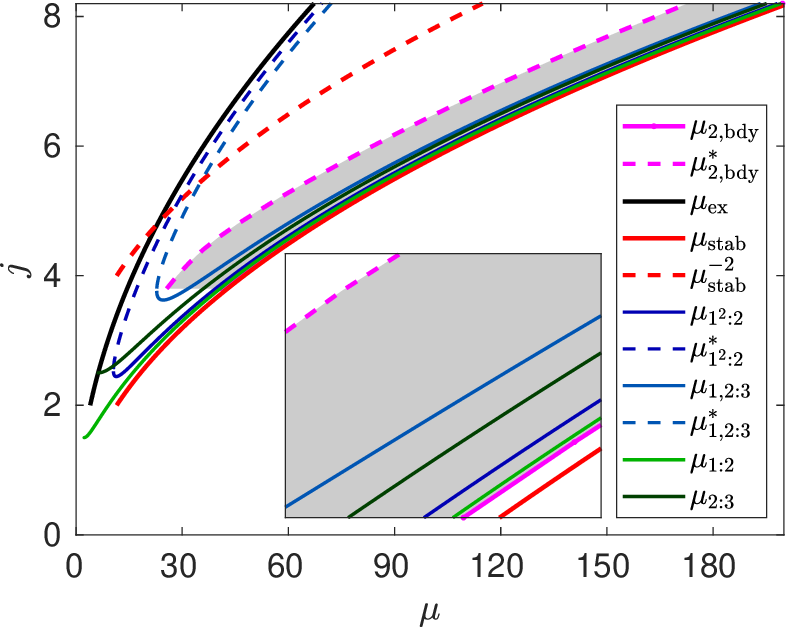} 
    \caption{Testing the linear prediction hypothesis in the fixed-$\mu$ problem, we computed trajectories with small perturbations of an unstable  equilibrium $E_j$ in the modes $\ell=1$ (left) and $\ell=2$ (right). We found that the trajectories converged to the equilibrium $E_{j-\ell}$ in a range (shaded gray) far exceeding the possible drop ranges investigated below. In the left figure, $\ell=1$, drops to $E_{j-1}$ were confirmed up for all $\mu$ larger than and up to a critical value $\mu_{1,\mathrm{bdy}}$, which includes a region where $\lambda_2>\lambda_1$ (shown as $\mu>\mu_{1:2}$) and a region where $\lambda_2>2\lambda_1$ (shown as $\mu>\mu_{1^2:2}$). However, it does not encompass the entire parameter region where $E_{j-\ell}$ is stable (shown as $\mu>\mu_\mathrm{stab}^{-1}$).  Remarkably, higher drops appear to occur past a somewhat fixed distance to the instability boundary $\mu_\mathrm{stab}$. On the right, the region with consistent drop-by-2 (shaded  gray) also encompasses well the region where we predict a local drop-by-two.  We refer to~\eqref{e:ress} for precise definitions of the various resonance curves shown and the main prediction in \S\ref{ss:main} relying on these resonances to predict drop numbers and drop times.  }\label{fig:hypothesis 1 evidence}
\end{figure}

Hypothesis~\ref{hyp: mode conservation} is reasonably well supported by numerical experiments; see Fig.~\ref{fig:hypothesis 1 evidence} for experiments in the cases $\ell=1$ and $\ell=2$. In the numerical experiments, we perturbed the unstable equilibrium in the eigenvector to $\lambda_{\ell}$, only, thus checking the milder version of the hypothesis with $\delta_1=0$. We found consistent drops by $\ell$ in a large region (shaded gray in the figure). We superimposed this region with eigenvalue resonance curves that we computed using numerical continuation. In the left panel, we show the transition from drop-by-one to drop-by-two as predicted in the next section at $\mu_{1^2:2}$. This resonance occurs well inside the shaded gray region. Note that for smaller $\mu$ the resonance is not present, the numerical continuation of the resonance condition undergoes a saddle-node with a second resonance curve $\mu^*_{1^2:2}$. In the right panel, we show in particular the resonance $\mu_{1,2:3}$, which indicates the transition to a drop-by-three, again well inside the region where the hypothesis holds for $\ell=2$; for details, compare the definition of the log amplitudes $b_j$~\eqref{e:b1}--\eqref{e:b4} and the main prediction~\eqref{e:droppred}.

In the spirit of this assumption, we fix $\delta>0$ sufficiently small and consider a trajectory that starts in a neighborhood of the unstable equilibrium $E_j$. We define the \emph{local drop number} (depending on $\delta$) at the time when the distance to the $E_j$ reaches $\delta$ by finding the index of the unstable eigenspace with maximal amplitude  in the sense of Hypothesis~\ref{hyp: mode conservation}. In the results presented in the next section, it then turns out that the distance to the eigenspace is in fact very small, that is, $\delta_1\ll\delta_0$.

We say that the  \emph{global drop number} of such a trajectory is $\ell$ if $E_{j-\ell}$ is the first equilibrium so that  the trajectory reaches its neighborhood of size $\delta$.
Hypothesis~\ref{hyp: mode conservation} is geared to imply that local drop numbers and global drop numbers are equal in the static ($\mu$ fixed) problem. In fact, local and global drop numbers agree well also in the dynamical problem, where $\mu$ varies slowly; compare Fig.~\ref{f:hd}.


We note that, thinking about proving a version of Hypothesis~\ref{hyp: mode conservation}, detailed information on connecting orbits usually requires information in addition to energy levels of equilibria or more general topological information ~\cite{MR1347417}. One such property are comparison principles with the  associated refined Sturm oscillation properties in the real scalar version of~\eqref{e: GL}, known as the Allen-Cahn equation, $u_t = u_{xx} + \mu u - u^3$. Here, in fact the  structure of connecting orbits is completely determined; see for instance~\cite[Chapter 5.3]{Henry}, the review~\cite{FS}, or~\cite{FR} for a more recent perspective. Some extensions towards systems of the type considered here were investigated in~\cite{buger}.


\section{Slow passage through the Eckhaus instability}
Based on our understanding of the static problem, in particular Hypothesis~\ref{hyp: mode conservation}, we now turn to predictions for the dynamic problem
\begin{align}
    B_t &= (\partial_x+\rmi j)^2 B + \mu B - B|B|^2, \label{e: dynamic system 1}\\
    \dot{\mu} &= - \eps. \label{e: dynamic system 2}
\end{align}
Throughout, we envision an initial condition given as a small generic perturbation of  the $j$-modal equilibrium $E_j$, for a $\mu$-value initially large enough such that $E_j$ is stable. As time evolves, the solution will follow the $\mu$-dependent equilibrium very closely as long as $\mu>\mu_\mathrm{c}$. Our goal is to predict when (i.e., for which $\mu$-value less than $\mu_\mathrm{c}$) the solution will leave a small neighborhood of $E_j$ and what the dominant linear mode $\ell$ is at that time instance, hence predicting a drop-by-$\ell$ based on our linear mode selection assumption, Hypothesis~\ref{hyp: mode conservation}.

\subsection{Main result -- spatio-temporal resonances and drop criteria}\label{ss:main}

We summarize here the results of the analysis in the subsequent sections in the form of an algorithm that predicts, at leading order, local drop parameter values $\mu_\mathrm{out}$ and local drop number $m$ given an initial parameter value $\mu_\mathrm{in}$ and initial wavenumber $j$ in~\eqref{e: dynamic system 2}. Recall the eigenvalues of the linearization $\lambda_{\ell}(\mu)$ and define resonant parameter values $\mu=\mu_{k_1,\ldots,k_s:k_0}$  as the values of $\mu$ where a resonance condition in the eigenvalues holds:
\begin{align}
\lambda_{k_0}(\mu)&=\sum_{\sigma=1}^s \lambda_{k_\sigma}(\mu) \quad \text{ at } \quad  \mu=\mu_{k_1,\ldots,k_s:k_0},\label{e:rest}\\
k_0&=\sum_{\sigma=1}^s k_\sigma.\label{e:ress}
\end{align}
For instance,
\[
\lambda_{3}(\mu)=\lambda_{1}(\mu)+\lambda_{2}(\mu),\quad \text{ at } \quad \mu=\mu_{1,2:3},
\]
and, introducing short-hand notation,
\[
\lambda_{3}(\mu)=3\lambda_{1}(\mu)\quad \text{ at } \quad \mu=\mu_{1,1,1:3}=:\mu_{1^3:3}.
\]
The parameter values $\mu_{k_1,\ldots,k_s:k_0}$ represent \emph{spatio-temporal resonances} in the sense that at these parameter values linear solutions $\rme^{\rmi kx+\lambda t}$ exist that are resonant both in time~\eqref{e:rest} and space~\eqref{e:ress}.

Next, we recursively define $\log$-amplitudes of modes $b_\ell$ given $\mu_\mathrm{in}$ as follows:
\begin{align}
b_1(\mu)&=\int_{\mu}^{\mu_\mathrm{in}}\lambda_1(\hat\mu)\rmd\hat\mu,\label{e:b1}\\
b_2(\mu)&=\int_{\mu}^{\mu_{1^2:2}}\lambda_2(\hat\mu)\rmd\hat\mu + 
                       2 b_1(\mu_{1^2:2}), \qquad \mu<\mu_{1^2:2}\label{e:b2}\\
b_3(\mu)&=\int_{\mu}^{\mu_{1,2:3}}\lambda_3(\hat\mu)\rmd\hat\mu + 
                        b_1(\mu_{1,2:3})+b_2(\mu_{1,2:3}), \qquad \mu<\mu_{1,2:3}\label{e:b3}\\
b_4(\mu)&=\int_{\mu}^{\mu_{2,2:4}}\lambda_4(\hat\mu)\rmd\hat\mu + 
                        2b_2(\mu_{2^2:4}) \qquad \mu<\mu_{2,2:4}.\label{e:b4}
\end{align}
\textbf{Main prediction --- drops $m\leq 4$.} \emph{Fix $\ell$ and an initial parameter value $\mu_\mathrm{in}>\mu_\mathrm{c}$, and consider dynamics with slowly decreasing $\mu=\mu_\mathrm{in}-\eps t$. The drop parameter value $\mu_\mathrm{out}$ is given to zeroth  order in $\eps$ by
\begin{equation}\label{e:mupred}
    \mu_{\mathrm{out}}=\mathrm{argmax}_{\mu<\mu_\mathrm{c}} \mathrm{max}_{1\leq \ell\leq 4 } b_\ell(\mu)=0,
\end{equation}
that is, the largest (dynamically the first) $\mu$-value at which any of the log amplitudes return to 0. The local drop number is
\begin{equation}\label{e:droppred}
    m= \mathrm{argmax}_{1\leq\ell}\, b_\ell(\mu_{\mathrm{out}}),
\end{equation}
that is, the index $\ell$ that achieves $b_\ell(\mu_{\mathrm{out}})=0.$ The maxima are taken only over log amplitudes that are defined for the given parameter value $\mu$. As a function of $\mu_\mathrm{in}$, $\mu_\mathrm{out}<\mu_\mathrm{c}$ is strictly decreasing and $m$ is strictly increasing. }

The $\log$-amplitudes before the resonances can be defined as well, but they are always much smaller than the $\log$-amplitudes with smaller index. 

The reader will notice that only specific resonances occur in the definition of the $b_\ell$. We found these resonances to be relevant in the sense that their effects dominate the effects of all other resonances in the examples we studied, in particular in the limit of large $j$. We do not know of a general rule to predict which resonances dominate for larger potential drop numbers, so the algorithm for drops by $5$ or higher is more complex. For the mode $b_5$, one needs to find $\log$-amplitudes depending on different spatio-temporal resonances and maximize over all of those, for instance defining
\begin{align*}
b_5^{2,3:5}(\mu)&=\int_{\mu}^{\mu_{2,3:5}}\lambda_5(\hat\mu)\rmd\hat\mu + 
                        b_2(\mu_{2,3:5}) + 
                        b_3(\mu_{2,3:5}) \qquad \mu<\mu_{2,3:5},\\
b_5^{1,4:5}(\mu)&=\int_{\mu}^{\mu_{1,4:5}}\lambda_5(\hat\mu)\rmd\hat\mu + 
                        b_1(\mu_{1,4:5}) + 
                        b_4(\mu_{1,4:5}) \qquad \mu<\mu_{1,4:5},\\
b_5^{1,2^2:5}(\mu)&=\int_{\mu}^{\mu_{1,2^2:5}}\lambda_5(\hat\mu)\rmd\hat\mu + 
                        b_1(\mu_{1,2^2:5}) + 
                        2b_2(\mu_{1,2^2:5}) \qquad \mu<\mu_{1,2^2:5},\qquad \text{etc.}
\end{align*}
With this information, we can define critical parameter values  where changes in drop numbers occur. At these parameter values, the argmax in~\eqref{e:droppred} is realized simultaneously at two different values of $\ell$. In all situations we encountered, those two values of $\ell$ were adjacent, $\ell\in\{m,m+1\}$ and the drop number increases by one as $\mu_\mathrm{in}$ increases through this critical parameter value. We then denote this critical parameter value by $\mu_{\mathrm{in},m\to m+1}$, with associated local drop parameter values $\mu_{\mathrm{out},m\to m+1}$.

It turns out that  $\mu_{\mathrm{out},1\to 2}=\mu_{1^2:2}$, but for $m\geq 2$, $\mu_{\mathrm{out},m\to m+1}$ is strictly less than the resonances associated with the mode $m+1$, for instance $\mu_{1,m:m+1}$, $\mu_{1^2,m-1:m+1}$, etc. We will see that, as a consequence, corrections to the transition points are $\rmO(\sqrt{\eps})$ for $\mu_{\mathrm{out},1\to 2}$ but $\rmO(\eps)$ for other transitions.

We refer to Figs.~\ref{f:1:2},~\ref{f:2:3}, and~\ref{f:hd} for numerical demonstration of the validity of the results presented here. The remainder of this section presents the rationale behind these predictions, explaining to what extent one should expect rigorous results, and some details on $\eps$-corrections as well as asymptotics for $j\to\infty$. 
The numerical results also show that global drop numbers agree well with local drop number predictions. This is also confirmed by the direct testing of Hypothesis~\ref{hyp: mode conservation}, Fig.~\ref{fig:hypothesis 1 evidence}, which shows that the Hypothesis holds well beyond the region where drop number changes are expected, at least for drop numbers of 1 and 2.



\subsection{Drop-by-1 transitions} 
First, we consider initial conditions for which $B_0$ is  close to an equilibrium $E_j$, and the initial parameter value $\mu_\mathrm{in}$ is just above the stability boundary $\mu_\mathrm{c}=\mu_{1,\mathrm{c}}$ at which $\lambda_{1}(\mu)$ becomes positive. Starting very close to the stability boundary, we expect the first drop in the winding number  to occur while $\lambda_{1}(\mu)$ is the only unstable eigenvalue, and so we study the reduced problem on the associated center manifold. In this case, no further assumptions are necessary to make precise and rigorous statements about the local drop time.

When $\eps = 0$ and $\mu =\mu_{1,\mathrm{c}}$, the static problem has a complex one-dimensional, real two-dimensional center manifold. A parameter-dependent center manifold reduction~\cite{BarkleyTuckerman} gives the leading order equation for the complex amplitude $a_1$ near $\mu = \mu_{1,\mathrm{c}}$,
\begin{align}
    \dot{a}_1 = \lambda_{1} (\mu) a_1 + c_1(\mu) a_1 |a_1|^2, \quad c_1(\mu) > 0,
\end{align}
which is the normal form for a sub-critical pitchfork bifurcation with rotational symmetry. The cubic coefficient, obtained through a center-manifold expansion was shown to be positive in~\cite{BarkleyTuckerman}. Solutions to~\eqref{e: GL conj static} on this center manifold are then recovered using the eigenvectors $e_1(\mu)$~\eqref{e:lamlj} through
\begin{align}
    B(x,t) = B_*(j) + u +  \rmi v,\qquad 
    u +  \rmi v = a_1(t) e_{1}(\mu) \rme^{\rmi x} + \bar{a}_1(t) \bar{e}_{1}(\mu) \rme^{-\rmi  x} + \mathrm{O}\left(|a_1|^2\right). \label{e:1mode}
\end{align}

Considering these center manifolds in the system~\eqref{e: dynamic system 1}--\eqref{e: dynamic system 2} for $\eps = 0$, we find an invariant manifold, of real dimension three, given by the union over $\mu$ of the center manifolds constructed above. This manifold is normally hyperbolic from the explicit splitting in the linearization at the equilibria and hence persists for $\eps$ small~\cite{Henry,MR1445489}, with dynamics given by
\begin{align}
    \dot{a}_1 &= \lambda_{1} (\mu) a_1 + c_1 (\mu) a_1 |a_1|^2+\mathrm{O}\left(|a_1|^5\right), \label{e: 1 drop slow manifold 1} \\
    \dot{\mu} &= - \eps. \label{e: 1 drop slow manifold 2}
\end{align}
As noted above, the flow commutes with the action of translations $a_1\mapsto \rme^{\rmi\varphi}a_1$ and reflections $a_1\mapsto \bar{a}_1$, so that the real subspace is invariant and dynamics for arbitrary initial conditions are conjugate to the dynamics in the real space by complex rotation. 
Alternatively, one could also use a 
time-dependent center-manifold reduction~\cite{MR1488358} after an appropriate modification of the $\mu$-dynamics outside of the parameter regime considered and arrive at the same reduced equation.  

We then consider a  boundary-value problem where we prescribe the initial amplitude $a_{1,\mathrm{in}}=\delta\ll 1$ and initial parameter value $\mu_\mathrm{in}$ at a time $-T_\mathrm{in}$, and seek time $T_\mathrm{out}$ and associated parameter value $\mu_\mathrm{out}$ where the amplitude $a_1$ is of size $\delta$ again,
\begin{align}
a_1(-T_\mathrm{in}) &= a_{1,\mathrm{in}}=\delta,\qquad\mu (-T_\mathrm{in}) = \mu_{\mathrm{in}} \nonumber \\
 a_1(T_\mathrm{out}) &= a_{1,\mathrm{out}} = \delta,\qquad \mu (T_\mathrm{out}) = \mu_{\mathrm{out}}. 
\label{e:bvp1}
\end{align}
For convenience, we set
\begin{equation}
\mu_{\mathrm{in}}-\mu_{1,\mathrm{c}}  = \eps T_\mathrm{in}, \text{ so that }
\mu_{1,\mathrm{c}}-\mu_{\mathrm{out}}  = \eps T_\mathrm{out},\quad
\mu(0)  =\mu_{1,\mathrm{c}}.\label{e:1bc}
\end{equation}
Because $\lambda_{1}(\mu_\mathrm{in})$ is stable, we expect that $a_1$ will initially decay exponentially, and then grow once $\mu$ decreases past $\mu_{1,\mathrm{c}}$, so  that the time $T_\mathrm{out}$ and thereby the value of $\mu_\mathrm{out}$ at which $a_1$ returns to its initial size, $a_1(T_\mathrm{out}) = \delta$, are well defined.

Since we are only considering a time interval in which $a_1$ remains small, as a first approximation we analyze~\eqref{e: 1 drop slow manifold 1}--\eqref{e: 1 drop slow manifold 2} by  neglecting the nonlinear terms and solving the resulting linear equation explicitly for initial conditions $(a_{1,\mathrm{in}},\mu_\mathrm{in})$, with solution
\begin{align}
    a_1(t) = \exp \left( \int_{-T_\mathrm{in}}^t \lambda_{1}(\mu(\eps s)) \, \rmd s \right) a_{1,\mathrm{in}}. 
\end{align}
Since $\lambda_{1}(\mu_\mathrm{in}) < 0$, the integral is initially negative and begins increasing as $\mu$ decreases past $\mu_{1,\mathrm{c}}$. Then in this linear prediction, the time $T_\mathrm{out}$ is the first time after $T_\mathrm{in}$ at which the equal area condition
\begin{equation}
    \int_{-T_\mathrm{in}}^{T_\mathrm{out}} \lambda_{1}(\mu(\eps s)) \, \rmd s = 0, \qquad \text{ or, equivalently, }\quad
     \int_{\mu_\mathrm{in}}^{\mu_\mathrm{out}} \lambda_{1}(\mu) \, \rmd\mu  = 0,
    \label{e: equal area condition}
\end{equation}
holds. Note that the latter integral describes precisely the vanishing $\log$-amplitude $b_1=0$ from \S\ref{ss:main}.

\begin{lemma}[Equal-area prediction]\label{l:ea}
The system~\eqref{e:1mode} together with the boundary conditions~\eqref{e:1bc} has a unique solution for $0<\delta, \mu_{1,\mathrm{in}}-\mu_{1,\mathrm{c}}  \ll 1$,  sufficiently small, and we have the expansion
\begin{equation}\label{e:ea}
\mu_\mathrm{out}=\mu_\mathrm{out}^0+\rmO(\eps), 
\end{equation}
where $\mu_\mathrm{out}^0<\mu_\mathrm{in}$ is the largest solution to the equal-area condition~\eqref{e: equal area condition}. In particular, $\mu_{1,\mathrm{c}}-\mu_\mathrm{out}=\mu_\mathrm{in}-\mu_{1,\mathrm{c}}+\rmO(|\delta|+|\eps|+|\mu_\mathrm{in}-\mu_{1,\mathrm{c}}|^2)$.
\end{lemma}
\begin{Proof}
    The result states that at leading order, the linear prediction~\eqref{e: equal area condition} is correct. For small $\delta$, the eigenvalue $\lambda_{1}(\mu)$ can be approximated near $\mu_\mathrm{1,c}$ by $\lambda_{1}(\mu)=\lambda_{1}'(\mu_{1,\mathrm{c}})(\mu-\mu_{1,\mathrm{c}})+\rmO(2)$, which implies the last statement.

    In order to show that nonlinear terms contribute only at higher order, one follows the analysis of the slow passage near a pitchfork bifurcation in~\cite{ks2}. We omit the somewhat lengthy but straightforward details. 
\end{Proof}
\begin{corollary}[drop-by-1 result]\label{c:1}
Assume Hypothesis~\ref{hyp: mode conservation} and an initial condition in a sufficiently small neighborhood to the $j$-mode equilibrium for some $j>0$, a parameter-value $\mu_\mathrm{in}\gtrsim \mu_\mathrm{1,c}$, and let $\eps$ be sufficiently small. Then the modal number will drop by one at $\mu_\mathrm{out}=2\mu_\mathrm{1,c}-\mu_\mathrm{in}+\rmO(\eps)$.
\end{corollary}
\begin{Proof}
The result follows immediately from Lemma~\ref{l:ea} and Hypothesis~\ref{hyp: mode conservation}.
\end{Proof}
From Corollary~\ref{c:1}, we conclude that for $\mu_\mathrm{in}$ sufficiently close to $\mu_{1,\mathrm{c}}$, we will always observe a drop by one and can easily characterize the drop time. Naturally, we wish to see how far this result extends. The key limiting factor in the result is the presence of a leading eigenvalue $\lambda_{1}(\mu)>\lambda(\mu)$ for all other $\lambda(\mu)$ in the spectrum of the linearization. Technically, center manifold reductions in the literature also require a \emph{uniform} splitting, $\lambda_{1}(\mu)>\eta_0>\lambda(\mu)$ with $\eta_0$ independent of $\mu$, possibly with sufficiently large gaps to ensure smoothness. We believe that a simple splitting of the leading eigenvalue $\lambda_{1}(\mu)>\lambda(\mu)$ for all $\mu\in (\mu_\mathrm{in},\mu_\mathrm{out})$ should be sufficient to establish Lemma~\ref{l:ea} but will not attempt to do so here. We observed excellent validity of our predictions even when the uniform gap condition does not hold.

The procedure thus far predicts how and when a first transition will lead to a drop by one in wavenumber or possibly a more severe drop. At the end of this drop, from say $E_j$ to $E_{j-1}$ near
$\mu=\mu_\mathrm{out}$, we have followed a global heteroclinic to a neighborhood of $E_{j-1}$, where we typically expect that upon entering the neighborhood, we have a generic perturbation of this stable equilibrium and can now repeat the analysis with the new $\mu_\mathrm{in}$ given by the previous $\mu_\mathrm{out}$. A short calculation using our main prediction~\eqref{e:droppred} and expressions for eigenvalues in~\eqref{e:lamlj}, shows that for $j\gg 1$, the distance between $\mu_{1,\mathrm{c}}$ for modes $j$ and $j+1$ is $6j+1$, while the distance between  $\mu_{1,\mathrm{c}}$ and  $\mu_{2,\mathrm{c}}$ is only $\frac{3}{2}$. Hence one expects that, for large $j$, a drop-by-1 is followed by a large distance to criticality in parameter space and a potentially subsequent higher drop; compare also Fig.~\ref{f:staircase}.
Therefore, we now analyze such higher drops, when more than one eigenvalue is unstable at the predicted drop time. 


\subsection{The drop-by-2 transition}\label{s: drop by two}
We first consider the static problem with $\mu_{3,\mathrm{c}} < \mu < \mu_\mathrm{2,c}$, so that the equilibrium $E_j$ has two unstable eigenvalues $\lambda_{1}(\mu)$ and $\lambda_{2}(\mu)$. There is an associated complex two-dimensional, real four-dimensional center-unstable manifold. The dynamics on this manifold are  governed by equations
for the complex amplitudes $a_1, a_2$ in the associated eigenspaces, 
\begin{align}
    \dot{a}_1 &= \lambda_{1}(\mu) a_1 + c_{1,2} (\mu) \bar{a}_1 a_2 +\rmO(|a_1|(|a_1|^2+|a_2|^2)),\nonumber \\
    \dot{a}_2 &= \lambda_{2} (\mu) a_2 + c_{1^2} (\mu) a_1^2 +\rmO(|a_1|(|a_1|^2+|a_2|^2)),\label{e:12}
\end{align}
with coefficients $c_{1,2}(\mu), c_{1^2}(\mu) \in \C$ that can be readily computed by evaluating the nonlinearity on the associated eigenspace and projecting.  Again, the vector field commutes with the isotropy of the equilibrium, that is, translations $(a_1,a_2)\mapsto (\rme^{\rmi\varphi} a_1,\rme^{2\rmi\varphi} a_2),$ and reflections $(a_1,a_2)\mapsto (\bar{a}_1,\bar{a}_2)$. As  consequence, $c_{1^2}$ and $c_{1,2}$ are real and the real subspace $(a_1,a_2)\in\R^2$ is invariant. Note that the nonlinear terms correspond to the spatial resonances referred to in~\eqref{e:ress}: the $a_j$ are amplitudes of Fourier modes $\rme^{\rmi j x}$, and the complex rotation is induced by the spatial shift $x\mapsto x+\varphi$ in the full equation.The associated manifold is normally hyperbolic in parameter regimes where $\lambda_{1}(\mu)$ and $\lambda_{2}(\mu)$ are separated from other eigenvalues; its smoothness (making in particular the cubic terms meaningful) is determined by spectral gaps; see for instance~\cite{Henry,hps}. These gaps can be easily evaluated in the large-$j$ limit~\eqref{e:lamlj} where $|\lambda_{2}(\mu)|/|\lambda_{3}(\mu)|>2$ for all $\hat{\mu}>0$, providing the desired smoothness of the manifold and validity of expansions.

For $\eps \neq 0$, this manifold persists as a slow manifold  with  dynamics at quadratic order given by
\begin{align}
     \dot{a}_1 &= \lambda_{1}(\mu) a_1 + c_{1,2} (\mu) \bar{a}_1 a_2 , \label{e: 2 drop slow manifold 1}\\
    \dot{a}_2 &= \lambda_{2} (\mu) a_2 + c_{1^2} (\mu) a_1^2, \label{e: 2 drop slow manifold 2}\\
    \dot{\mu} &= - \eps. \label{e: 2 drop slow manifold 3} 
\end{align}
%
To predict drop times and drop numbers, we will try to determine when $(a_1,a_2)$ leave a small neighborhood of the origin, and whether $|a_1|\gg |a_2|$ or $|a_2|\gg |a_1|$ at that time, thus concluding that the \emph{local drop number} is $1$ or $2$, respectively. Hypothesis~\ref{hyp: mode conservation} then allows us to conclude global drop numbers. We therefore consider the system~\eqref{e: 2 drop slow manifold 1}--\eqref{e: 2 drop slow manifold 3} with initial conditions $a_1(-T_\mathrm{in}) = a_2 (-T_\mathrm{in}) = \delta$, $\mu (-T_\mathrm{in}) = \mu_\mathrm{in}$, with $\mu_\mathrm{in} > \mu_{1,\mathrm{c}} > \mu_{2,\mathrm{c}}$ so that both eigenvalues are initially stable. As we shall quickly see, the initial value of $a_2$ is irrelevant so that this particular choice is not restrictive. The solution will therefore initially contract, and we look for the first time $T_\mathrm{out}$ at which
\begin{align}
    \max \left\{ | a_1(T_\mathrm{out} )|, |a_2 (T_\mathrm{out} )|  \right\} = \delta. 
\end{align}

\noindent \textbf{Asymptotics from variation-of-constant formula.} We neglect higher-order terms and study~\eqref{e: 2 drop slow manifold 1}--\eqref{e: 2 drop slow manifold 3} setting $c_{1,2}\equiv 0$, noticing that $a_2\ll1$ so that $|c_{1,2}\bar{a_1}a_2|\ll |a_1|$; see also~\eqref{e:proj12} and the discussion there. We also use $\mu$ as an equivalent time variable and find amplitudes
\begin{align}
    a_1(\mu)&=\rme^{\frac{1}{\eps}\int_{\mu}^{\mu_\mathrm{in}}\lambda_1(\nu)\rmd\nu}\delta,\nonumber\\
    a_2(\mu)&=a_2^0(\mu)+a_2^{1^2}(\mu),\nonumber\\
    a_2^0(\mu)&=\rme^{\frac{1}{\eps}\int_{\mu}^{\mu_\mathrm{in}}\lambda_2(\nu)\rmd\nu}\delta,\nonumber\\
    a_2^{1^2}(\mu)&=\int_{\mu}^{\mu_\mathrm{in}}c_{1^2}(\nu)\rme^{\frac{1}{\eps}\Lambda(\nu)}\rmd\nu\,\delta^2,\nonumber\\
    \Lambda(\nu;\mu)&=\int_\mu^\nu \lambda_2(\rho)\rmd\rho + \int_\nu^{\mu_\mathrm{in}}2\lambda_1(\rho)\rmd\rho.\label{e:a21120}
\end{align}
Here, we suppress the dependence on $\mu_\mathrm{in}$ in  $\Lambda(\nu;\mu)$ and notice that $\Lambda(\nu;\mu)$ is smooth and maximal when $\nu=\mu_{1^2:2}$ so that we can expand the integrand near $\nu=\mu_{1^2:2}$ to find 
\begin{align}
 a_2^{1^2}(\mu)&=c_{1^2}\rme^{\frac{1}{\eps}\Lambda}\delta^2 \int_{\mu}^{\mu_\mathrm{in}}\rme^{\frac{1}{2\eps} \Lambda_{\nu\nu}(\nu-\mu_{1^2:2})^2}\rmd\nu (1+\rmO(\sqrt\eps))\nonumber\\
 &=c_{1^2}\rme^{\frac{1}{\eps}\Lambda}\delta^2\sqrt{\frac{-2\eps}{\Lambda_{\nu\nu}}}\mathrm{err}\,\left(\frac{\mu_{1^2:2}-\mu}{\sqrt{\eps}}\right)\left(1+\rmO(\sqrt\eps)\right),\label{e:a2112}
\end{align}
where $\Lambda_{\nu\nu}=(\lambda_2-2\lambda_1)'<0$ is the derivative in $\mu$ of the resonance condition, and $c_{1^2},\Lambda$, and $\Lambda_{\nu\nu}$ are evaluated at $\nu=\mu_{1^2:2}$, and $\mathrm{err}\,(z)=\int_{-\infty}^z \exp(-z^2)\rmd z$.

First, one sees that $a_2^0\ll a_2^{1^2}$. In order to identify the larges amplitude upon exiting a neighborhood, thus determining if the local drop number is 1 or 2, we therefore need to compare $a_1$ and $a_2\sim a_2^{1^2}$. The transition from a local drop number of 1 to a drop number 2 occurs when \begin{equation}
a_2^{1^2}=\delta,\quad \text{and }\quad a_1=\delta.\label{e:mucrit}
\end{equation}
Solving these two equations for $\mu_\mathrm{in}$ and $\mathrm{out}$ then identifies the transition parameter values  $\mu_{\mathrm{in},1\to2}$ and $\mu_{\mathrm{out},1\to2}$. In order to solve~\eqref{e:mucrit}, we use the expression for $a_1$ to find
$\int_{\mu_{\mathrm{out},1\to 2}}^{\mu_\mathrm{in}}\lambda_1(\nu)\rmd\nu =0$. Substituting this result into the expression for $\Lambda$ in~\eqref{e:a21120}, we obtain
\begin{equation}\label{e:lrr}
\Lambda(\mu_{1^2:2},\mu_{\mathrm{out},1\to 2})=\int_{\mu_{\mathrm{out},1\to 2}}^{\mu_\mathrm{1^2:2}}(\lambda_2(\rho)-2\lambda_1(\rho))\rmd\rho.
\end{equation}
We next assume that  $\Delta\mu=\mu_{1^2:2}-\mu_{\mathrm{out},1\to 2}>0$ is small so that we can Taylor expand
\[
\Lambda(\mu_{1^2:2},\mu_{\mathrm{out},1\to 2})=-\frac{1}{2}\Lambda_{\nu\nu}(\mu_{1^2:2},\mu_{1^2:2})(\Delta\mu)^2+\rmO((\Delta\mu)^3)
.\]
Inserting this into~\eqref{e:a2112} and partially solving for $\Delta\mu$, we find
\begin{equation}
\Delta\mu=\sqrt{\frac{2\eps}{\Lambda_{\nu\nu}}\log\left( c_{1^2}\delta\sqrt{-2\eps/(\Lambda_{\nu\nu})} \mathrm{err}\,(\Delta\mu/\sqrt{\eps})\right)}.
\end{equation}
Assuming that at leading order $\Delta\mu\sim \sqrt{-\eps\log\eps}$, we find $\mathrm{err}\,(\Delta\mu/\sqrt{\eps})\to\sqrt{\pi}$, so that, consistently, at leading order
\begin{equation}
\Delta\mu=\sqrt{\frac{2\eps}{\Lambda_{\nu\nu}}\log\left( c_{1^2}\delta\sqrt{-2 \pi\eps/(\Lambda_{\nu\nu})}\right)}.
\end{equation}
Away from the transition, the amplitude of $a_2(\mu)$ is given at leading order through
\begin{equation}\label{e:a2amp}
    a_2(\mu)=\left\{\begin{array}{ll} 
      \rmO(\eps a_1^2(\mu)), & \mu>\mu_{1^2:2},\\
      \exp\left({\frac{1}{\eps} \int_{\mu}^{\mu_{1^2:2}}\lambda_2(\rho)\rmd\rho +\frac{1}{\eps}  \int_{\mu_{1^2:2}}^{\mu_\mathrm{in}}2\lambda_1(\rho)\rmd\rho }\right),& \mu<\mu_{1^2:2},
    \end{array}
    \right.
\end{equation}
or, neglecting $\log\eps$-terms, 
\begin{equation}\label{e:b2amp}
b_2(\mu)=\eps \log(a_2(\mu))=\int_\mu^{\mu_\mathrm{in}} \Lambda_\mathrm{max}(\rho)\rmd\rho, \qquad \Lambda_\mathrm{max}(\rho)=\mathrm{max}\,\{\lambda_2(\rho),2\lambda_1(\rho)\}=\left\{\begin{array}{ll}2\lambda_1(\rho),& \rho>\mu_{1^2:2},\\
\lambda_2(\rho),& \rho<\mu_{1^2:2}.
\end{array}\right.
\end{equation}

\begin{figure}\centering
    \includegraphics[width=0.4\textwidth]{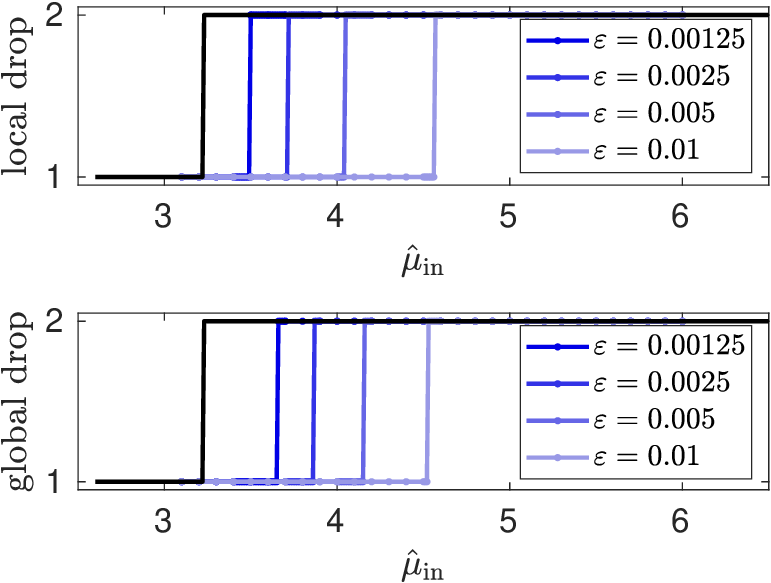}\qquad 
    \includegraphics[width=0.4\textwidth]{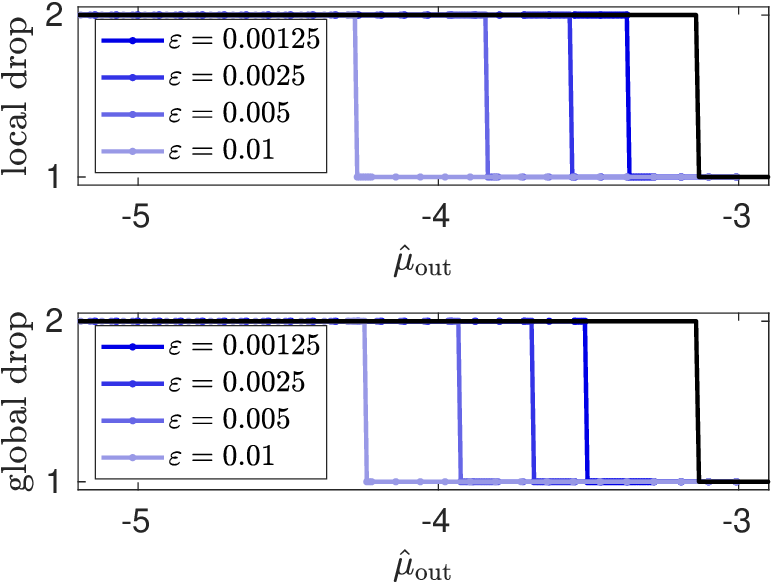}\\
    \includegraphics[width=0.4\textwidth]{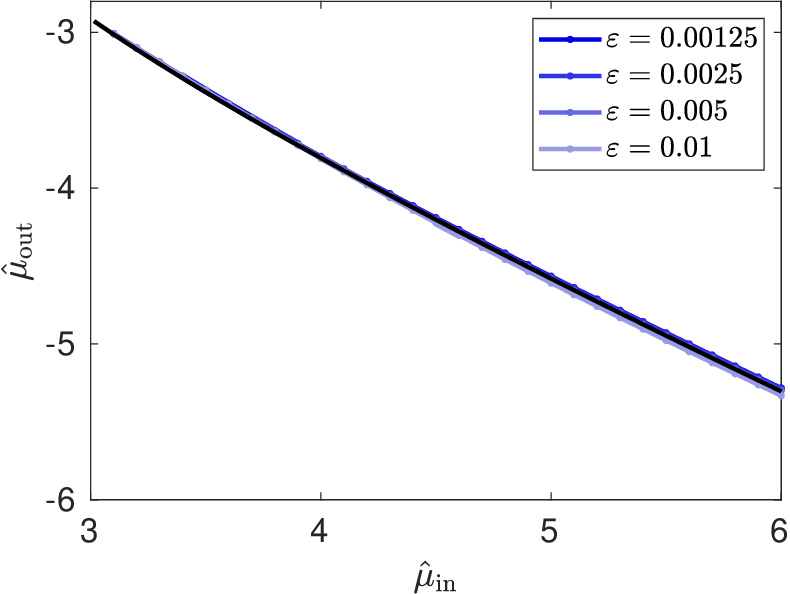}\qquad 
    \includegraphics[width=0.4\textwidth]{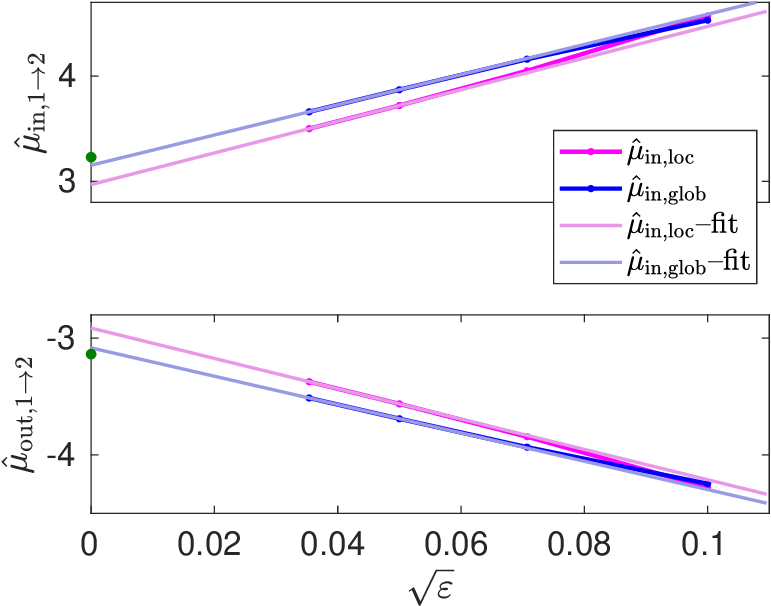} 
    \caption{Numerical comparisons for the drop-by-1 to  drop-by-2 transition. Top: Local and global drop numbers depending on $\mu_\mathrm{in}$ (left) and $\mu_\mathrm{out}$ (right) for several values of $\eps$; predicted drop as solid black curve. Bottom: $\mu_\mathrm{out}$ versus $\mu_\mathrm{in}$ for several values of $\eps$ together with prediction for $\eps=0$ (solid black) (left); drop values $\mu_\mathrm{in}$ both local and global as observed in top row shown here as a function of $\eps$, with linear fit and predicted values of drops as green dot. Extrapolated values at $\eps=0$ from linear fits fall within 10\% of the predicted value for local drops and within 2\% for global drops (right). Throughout, $\mu=\mu_\mathrm{c}+\hat{\mu}$ indicates values relative to criticality.}
    \label{f:1:2}
\end{figure}

For $\mu_\mathrm{out}<\mu_{\mathrm{out},1\to 2}$, at leading order the local drop parameter value is given by setting $a_2^{1^2}(\mu)=\delta$ in~\eqref{e:a21120}, which gives
\begin{equation}
\int_{\mu_\mathrm{out}}^{\mu_{1^2:2}}\lambda_2(\rho)\rmd\rho + \int_{\mu_{1^2:2}}^{\mu_\mathrm{in}}2\lambda_1(\rho)\rmd\rho \stackrel{!}{=}0,
\label{e:drop2time}
\end{equation}
in agreement with our predictions in \S\ref{ss:main} and with direct simulations illustrated in Fig.~\ref{f:1:2}.
%

\begin{figure}
    \centering
    \includegraphics[width=0.69\textwidth]{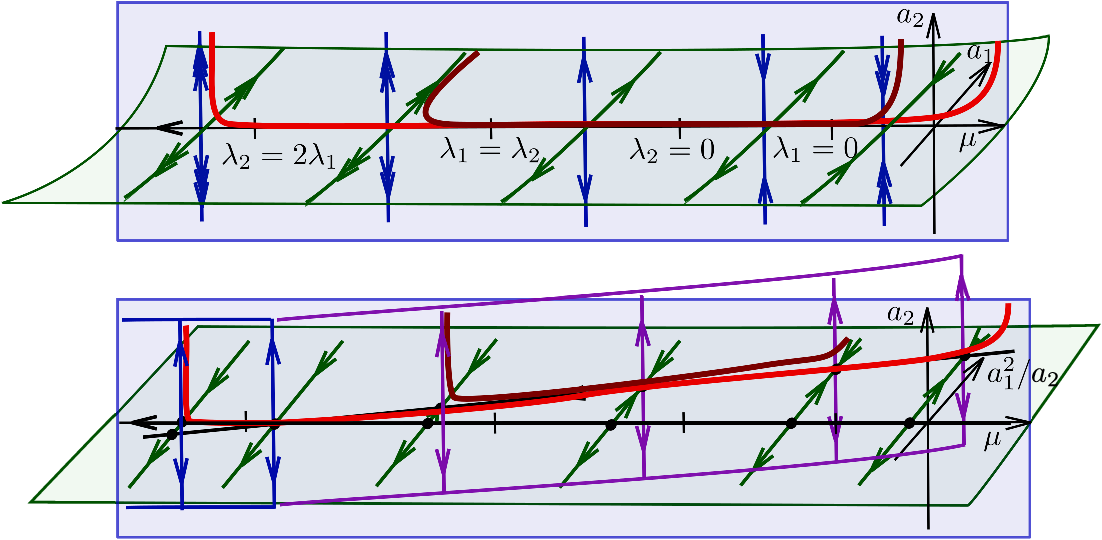}
    \caption{Top: Dynamics of center manifold equations~\eqref{e:12} with instability thresholds for $a_1$ and $a_2$ and $\lambda_{1/2}=0$, respectively, and subsequent 1:1 and 1:2 resonances; $\mu$ decreases to the left, the vertical plane $a_1=0$ is invariant. Sample dark and bright red trajectories illustrating trajectories exiting in the $a_1$ and $a_2$ directions, respectively. Bottom: dynamics in  projective coordinates~\eqref{e:proj12} with invariant plane $a_2=0$ and $\xi=a_1^2/a_2=0$; trajectories follow the stable branch in a transcritical bifurcation in the horizontal plane with unstable manifold in purple and blue for nontrivial and trivial branch, respectively; trajectories exit in the $a_2$ direction along the unstable manifold, albeit at locations where $a_1^2/a_2\neq 0$, so $|a_1|\sim \sqrt{|a_2|}\gg |a_2|$ (bright red), or when $a_1^2/a_2\sim 0$ and $|a_2|\gg |a_1|$ (dark red).}
    \label{f:geom}
\end{figure}

\paragraph{Rigorous asymptotics and a geometric view on resonances.} 
We next present a  geometric view that also explains why higher-order terms are irrelevant. Consider the new variable $\xi=a_1^2/a_2$, together with $a_2$, and find the new equations
\begin{align}
    \xi'=&(2\lambda_{1}(\mu)-\lambda_{2}(\mu))\xi -c_{1^2}(\mu)\xi^2 +\rmO(|a_2|)\nonumber,\\
    a_2'=&\lambda_{2}(\mu)a_2+c_{1^2}(\mu)\xi a_2+\rmO(|a_2|^2),\\
    \mu'=&\eps.\label{e:proj12}
\end{align}
In the invariant plane $a_2=0$, we find a slow passage through a transcritical bifurcation at the resonance, as $2\lambda_{1}(\mu)-\lambda_{2}(\mu)$ passes through zero. Before the resonance, $2\lambda_{1}(\mu)-\lambda_{2}(\mu)>0$, and the nontrivial equilibrium $\xi_*=(2\lambda_{1}(\mu)-\lambda_{2}(\mu))/c_{1^2}$ is exponentially attracting; past the resonance, trajectories follow the now stable trivial equilibrium $\xi=0$. Normal to this stable family of equilibria, trajectories first decay, then grow in the $a_2$-direction with the eigenvalue $\lambda_{2}+c_{1^2}\xi_*=2\lambda_{1}$; after the passage through the transcritical bifurcation growth is governed by the normal eigenvalue  $\lambda_{2}$, thus reproducing the integral criterion~\eqref{e:drop2time}. A refined desingularization, analyzing for instance the slow passage through the transcritical bifurcation using geometric blowup as in~\cite{ks2} should then lead to a leading-order expansion of the form obtained through the direct calculations outlined above; see Fig.~\ref{f:geom} for an illustration.

\subsection{The drop-by-3 transition}

Turning to predictions for higher drops, we need to account for more unstable eigenvalues and therefore higher resonances. We first adapt the asymptotic calculation for the 1:2-resonance criteria above to the drop-by-3 transition.

Once $\lambda_{3}$ becomes positive, we track the six-dimensional unstable manifold associated to the first three complex unstable modes in the static problem. Keeping track of what interactions of modes $\rme^{\pm 3ix}, \rme^{\pm 2 ix}, \rme^{\pm ix}$ can produce the original modes $\rme^{ix}, \rme^{2ix}, \rme^{3ix},$ we see that to leading order, the dynamics on this unstable manifold are given by
\begin{align}
    \dot{a}_1 &= \lambda_{1}(\mu) a_1 + c_{-1, 2} \bar{a}_1 a_2 + c_{3,-2} a_3 \bar{a}_2, \\
    \dot{a}_2 &= \lambda_{2}(\mu) a_2 + c_{1^2} a_1^2 + c_{3, -1} a_3 \bar{a}_1, \\
    \dot{a}_3 &= \lambda_{3} (\mu) a_3 + c_{1^3} a_1^3 + c_{1,2} a_1 a_2, \\
    \dot{\mu} &= - \eps, 
\end{align}
also taking into account the slow evolution of $\mu$. Error terms can be seen to be small either heuristically or in the analogue of the geometric desingularization shown in Figure~\ref{f:geom}; see the discussion below.
We again consider this system with initial conditions $a_1 (-\Tin) = a_2 (-\Tin) = a_3 (-\Tin) = \delta$, $\mu(-T_\mathrm{in})=\mu_\mathrm{in}$, and look for the first time $\Tout$ for which $|a_j(\Tout)| = \delta$ for some $j = 1,2$ or $3$. 

The coefficients of the nonlinear terms depend on $\mu$ and hence are also slowly varying, but since the values of these coefficients will not affect our predictions at leading order, we suppress this dependence. Also note that we have included the cubic term $c_{1^3}a_1^3$ in the equation for $a_3$, since the analysis of \S\ref{s: drop by two} suggests that for some time we have $a_2 \sim a_1^2$, and so $a_1^3$ terms are in fact on the same order as $a_1 a_2$ for some relevant time. 
Following a similar line of reasoning, we see as in \S\ref{s: drop by two} that the back-coupling terms involving $a_2$ and $a_3$ in the equation for $a_1$, and $a_3$ in the equation for $a_2$, are higher order. We therefore neglect these terms and arrive at the system
\begin{align}
    \dot{a}_1 &= \lambda_{1}(\mu) a_1, \label{e: 3 drop a1 eqn} \\
    \dot{a}_2 &= \lambda_{2}(\mu) a_2 + c_{1^2} a_1^2, \label{e: 3 drop a2 eqn} \\
    \dot{a}_3 &= \lambda_{3} (\mu) a_3 + c_{1^3} a_1^3 + c_{1,2} a_1 a_2, \label{e: 3 drop a3 eqn} \\
    \dot{\mu} &= - \eps. 
\end{align}
We can now solve this system recursively, first solving~\eqref{e: 3 drop a1 eqn} for $a_1$, then solving for $a_2$ via its variation of constants formula, and then inserting both expressions for $a_1$ and $a_2$ into~\eqref{e: 3 drop a3 eqn} and solving via the variation of constants formula. We shall use the expressions for $a_1$ and $a_2$ derived above and are left with the equation for $a_3$, which we write as a sum of the solution to the homogeneous equation $a_3^0$, the solution $a_3^{1^3}$ with inhomogeneity $c_{1^3}a_1^3$, and the solution $a_3^{1,2}$  with inhomogeneity $c_{1,2}a_1a_2$. We find
\begin{align}
    a_3^0(\mu)&=\rme^{\frac{1}{\eps}\int_{\mu}^{\mu_\mathrm{in}}\lambda_3(\nu)\rmd\nu}\delta,\nonumber\\
    a_3^{1^3}(\mu)&=\int_{\mu}^{\mu_\mathrm{in}}c_{1^3}(\mu)\rme^{\frac{1}{\eps}\Lambda_{{1^3}}(\nu)}\rmd\nu\delta^3,\nonumber\\
    \Lambda_{1^3}(\nu)&=\int_\mu^\nu \lambda_3(\rho)\rmd\rho + \int_\nu^{\mu_\mathrm{in}}3\lambda_1(\rho)\nonumber\\
    a_3^{1,2}(\mu)&=\int_{\mu}^{\mu_\mathrm{in}}c_{1,2}(\mu)\rme^{\frac{1}{\eps}\Lambda_{{1,2}}(\nu)}\rmd\nu\delta^3,\nonumber\\
    \Lambda_{1,2}(\nu)&=
    \int_\mu^\nu \lambda_3(\rho)\rmd\rho 
    + \int_\nu^{\mu_{1^2,2}}(\lambda_1(\rho)+\lambda_2(\rho))\rmd\rho
    + \int_{\mu_{1^2,2}}^{\mu_\mathrm{in}}3\lambda_1(\rho)\rmd\rho. \label{e:a21120_}
\end{align}
Following the same strategy as in the analysis for near the $1^2$:2-resonance, we evaluate the integrals to leading order near the maximum of the exponential, finding
\begin{align}
b_3^{1^3}&=\eps\log |a_3^{1^3}|=\Lambda_{1^3}(\mu_{1^3:3})+\rmO(\eps^{1^-}),\\
b_3^{1,2}&=\eps\log |a_3^{1,2}|=\Lambda_{1,2}(\mu_{1,2:3})+\rmO(\eps^{1^-}),
\end{align}
where we used the short hand $1^-$ to denote any exponent less than 1, accounting for potential terms of the form $\eps\log\eps$.
Clearly, $b_3^{1,2} > b_3^{1^3} > b_3^0$, so that the amplitude of $a_3$ is at leading order given through $\exp(b_3^{1,2}/\eps)$. The crossover occurs when $a_3\sim a_2$, that is, at leading order, when
\begin{align}
    &\int_{\mu_{\mathrm{out},2\to 3}}^{\mu_{1,2:3}} \lambda_3(\rho)\rmd\rho 
    + \int_{\mu_{1,2:3}}^{\mu_{1^2:2}}(\lambda_1(\rho)+\lambda_2(\rho))\rmd\rho
    + \int_{\mu_{1^2:2}}^{\mu_{\mathrm{in},2\to 3}}3\lambda_1(\rho)\rmd\rho \\
    \stackrel{!}{=} &\int_{\mu_{\mathrm{out},2\to 3}}^{\mu_{1^2:2}} \lambda_2(\rho)\rmd\rho 
    + \int_{\mu_{1^2,2}}^{\mu_{\mathrm{in},2\to 3}}2\lambda_1(\rho)\rmd\rho + \rmO(\eps^{1^-})\\
    \stackrel{!}{=} &\rmO(\eps^{1^-}).
\end{align}
\begin{figure}\centering
    \includegraphics[width=0.4\textwidth]{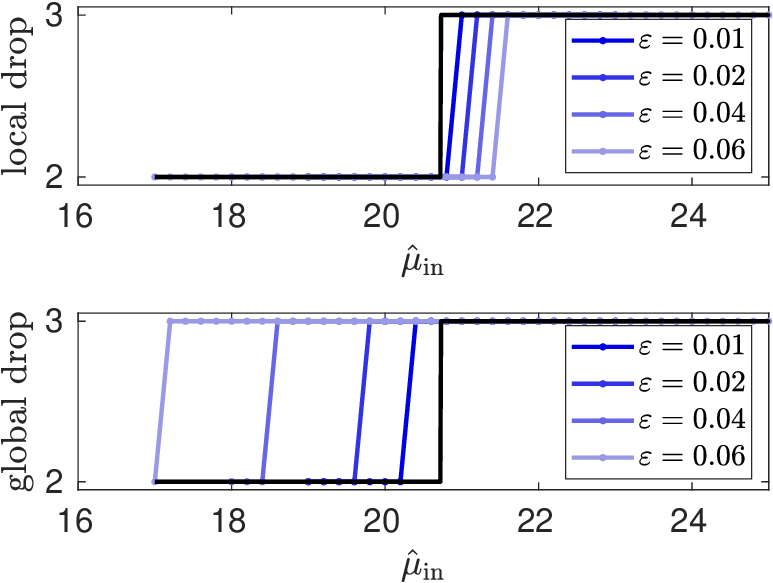}\qquad 
    \includegraphics[width=0.4\textwidth]{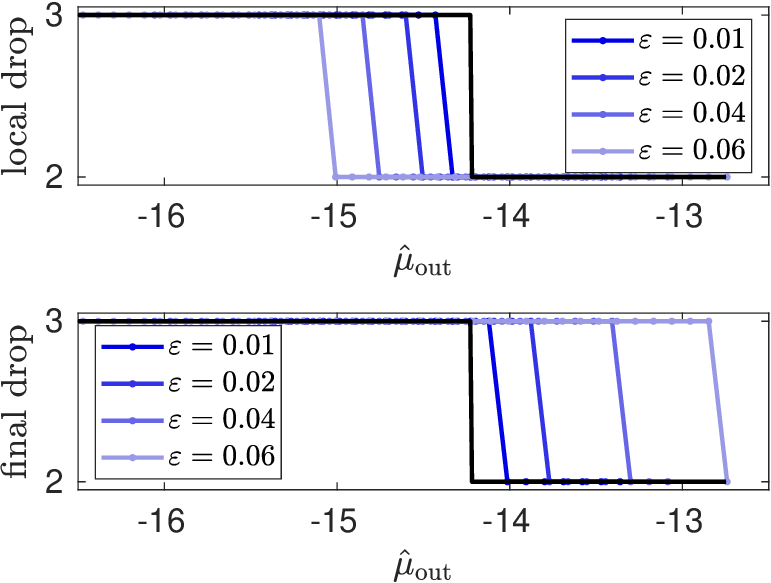}\\
    \includegraphics[width=0.4\textwidth]{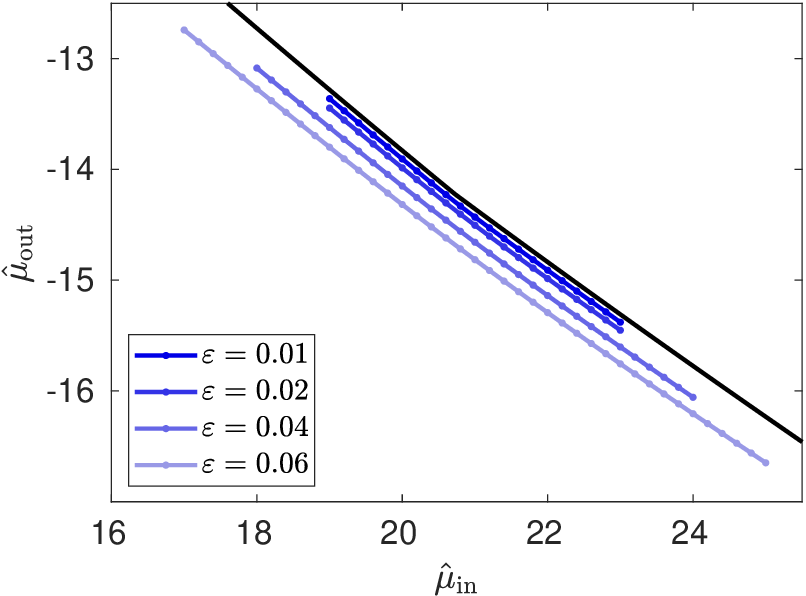}\qquad 
    \includegraphics[width=0.4\textwidth]{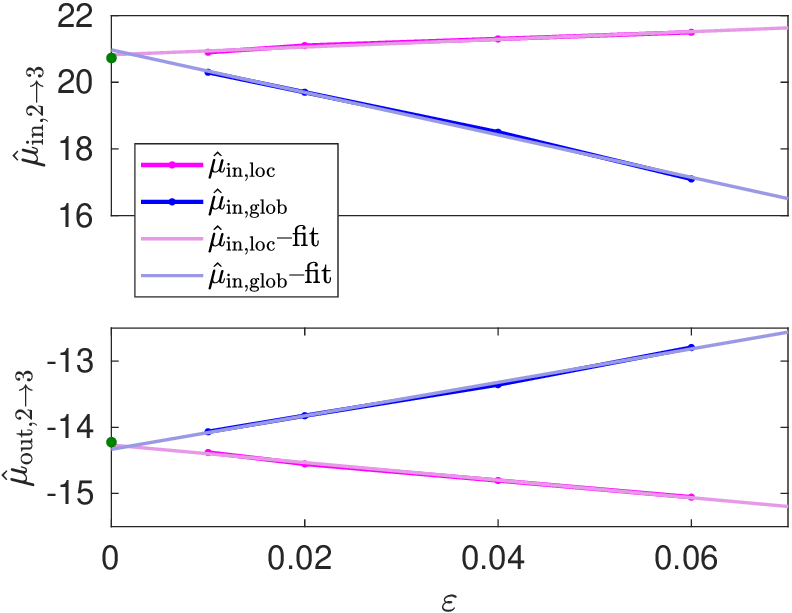} 
    \caption{Numerical comparisons for the drop-by-2 to drop-by-3 transition. Top: Local and global drop numbers depending on $\mu_\mathrm{in}$ (left) and $\mu_\mathrm{out}$ (right) for several values of $\eps$; predicted drop as solid black curve. Bottom: $\mu_\mathrm{out}$ versus $\mu_\mathrm{in}$ for several values of $\eps$ together with prediction for $\eps=0$ (solid black) (left); drop values $\mu_\mathrm{in}$ both local and global as observed in top row shown here as a function of $\eps$, with linear fit and predicted values of drops as green dot. Extrapolated values at $\eps=0$ from linear fits fall within 1\% of the predicted value (right). Throughout, $\mu=\mu_\mathrm{c}+\hat{\mu}$ indicates values relative to criticality.}\label{f:2:3}
\end{figure}

The two equalities determine $\mu_{\mathrm{out},2\to3}$ and $\mu_{\mathrm{in},2\to3}$. Linearizing at a solution with respect to these two variables, we find a Jacobi matrix with determinant 
\[
\left|\begin{array}{cc}
-\lambda_3(\mu_\mathrm{out})+\lambda_2(\mu_\mathrm{out})& 3\lambda_1(\mu_\mathrm{in})\\
-\lambda_2(\mu_\mathrm{out})& 2\lambda_1(\mu_\mathrm{in})
\end{array}\right|=\lambda_1(\mu_\mathrm{in})\left(-2 \lambda_3(\mu_\mathrm{out})+ 5 \lambda_2(\mu_\mathrm{out})\right).
\]
Assuming the additional non-resonance condition $2\lambda_3(\mu_\mathrm{out})\neq 5 \lambda_2(\mu_\mathrm{out})$ we therefore expect to be able to solve for the variables $\mu_\mathrm{in}$ and $\mu_\mathrm{out}$ with the Implicit Function Theorem with corrections to the transition value as  $\rmO(\eps^{1^-})$, rather than $\rmO(\sqrt{\eps})$ for the 1-2-transition.
The results agree with our summary in \S\ref{ss:main} and with numerical simulations shown in Fig.~\ref{f:2:3}.

\paragraph{The geometric picture for the drop-by-3 transition.}

We can follow the strategy for the analysis of the drop-by-2 transition and introduce projective variables that encode the quotients of amplitudes and resonances. For instance, set 
$
\xi_2=a_1^2/a_2, \xi_3=a_1a_2/a_3,
$ 
writing for short $\lambda_j=\lambda_{j}(\mu)$, suppressing $\mu$-dependence in the nonlinear terms, and suppressing higher-order terms in $a_1$, to find
\begin{align*}
    a_1'&=\lambda_1 a_1,\\
    \xi_2'&=(2\lambda_1-\lambda_2)\xi_2 -c_{1^2:2} \xi_2^2,\\
    \xi_3'&=(\lambda_1+\lambda_2-\lambda_3)\xi_3 - c_{1,2:3}\xi_3^2 - c_{1^3:3}\xi_3^2\xi_2.    
\end{align*}
We previously analyzed the dynamics in the $(a_2,\xi_2)$-subsystem. In the regime $\mu>\mu_{1^2:2}$, $\xi_3$ follows the nontrivial stable branch with $\xi_3=(\lambda_1+\lambda_2-\lambda_3)/(c_{1,2:3} - c_{1^3:3}\xi_2)$. For $\mu_{1,2:3}<\mu<\mu_{1^2:2}$, $\xi_2\sim 0$ and $\xi_3=(\lambda_1+\lambda_2-\lambda_3)/c_{1,2:3}$ becomes the nontrivial stable branch, which at $\mu_{1,2:3}$ exchanges stability with the trivial branch $\xi_3=0$ in a transcritical bifurcation. In particular, $\xi_3\sim 0$ for $\mu<\mu_{1,2:3}$. To reproduce the explicit computations above, recover growth by analyzing exponential growth in the normal direction of $a_1$, and of $a_{2/3}$ through the values of $\xi_{2,3}$. 

\subsection{Drop-by-4 and beyond}\label{s:hd}
Predicting higher drops appears to be cumbersome in general. We outline here the calculations for the transition from a drop-by-3 to a drop-by-4. The relevant amplitude equations are
\begin{align}
    \dot{a}_1 &= \lambda_{1}(\mu) a_1, \label{e: 4 drop a1 eqn} \\
    \dot{a}_2 &= \lambda_{2}(\mu) a_2 + c_{1^2} a_1^2, \label{e: 4 drop a2 eqn} \\
    \dot{a}_3 &= \lambda_{3} (\mu) a_3 + c_{1,2} a_1 a_2+ c_{1^3} a_1^3,  \label{e: 4 drop a3 eqn} \\
    \dot{a}_4 &= \lambda_{4} (\mu) a_4  + c_{1,3} a_1 a_3+ c_{2,2} a_2^2 + +c_{1^2,2}a_1^2a_2+ c_{1^4} a_1^4,\label{e: 4 drop a4 eqn} \\
    \dot{\mu} &= - \eps. 
\end{align}
Higher modes will be irrelevant until growth dominates the source terms from resonant interactions. In order to determine the onset of growth in the mode $a_4$, one would inspect the lowest order source terms, $a_2^2$ and $a_1a_3$, and find the associated resonances, $\mu_{2^2:4}$ and $\mu_{1,3:4}$, which give a lower bound for the crossover to a drop-by-4. 

It is convenient to track approximations to the logarithms of the amplitudes, $b_k=\eps\log a_k$, so that for instance
\begin{align}
    b_1(\mu)&=\int_\mu^{\mu_\mathrm{in}} \lambda_1(\mu)\rmd\mu,\nonumber\\
    b_2(\mu)&=\int_\mu^{\mu_{1^2:2}} \lambda_2(\mu)\rmd\mu+2b_1(\mu_{1^2:2}),\nonumber\\
    b_3(\mu)&=\int_\mu^{\mu_{1,2:3}} \lambda_3(\mu)\rmd\mu+b_1(\mu_{1,2:3}) +b_2(\mu_{1,2:3}).
\end{align}
For $b_4$, we need to consider all resonant terms in the equation for $a_4$. At leading order, we find that $b_4$ is given as a maximum of the expressions obtained by treating all terms separately,
\begin{align*}
    b_4^{1,3}(\mu)&=\int_\mu^{\mu_{1,3:4}} \lambda_4(\mu)\rmd\mu+b_1(\mu_{1,3:4})+b_3(\mu_{1,3:4}),\\
     b_4^{2,2}(\mu)&=\int_\mu^{\mu_{1,3:4}} \lambda_4(\mu)\rmd\mu+2b_2(\mu_{1,3:4}),\\
     \ldots&
\end{align*}
and similar expressions for cubic and quartic resonant terms. At any given $\mu<\mu_{1,3:4}$, the amplitude is given by the maximum $b_4=\max_m b_j^m$ where $m$ runs through all resonant source terms. 

\begin{figure}
    \includegraphics[width=\textwidth]{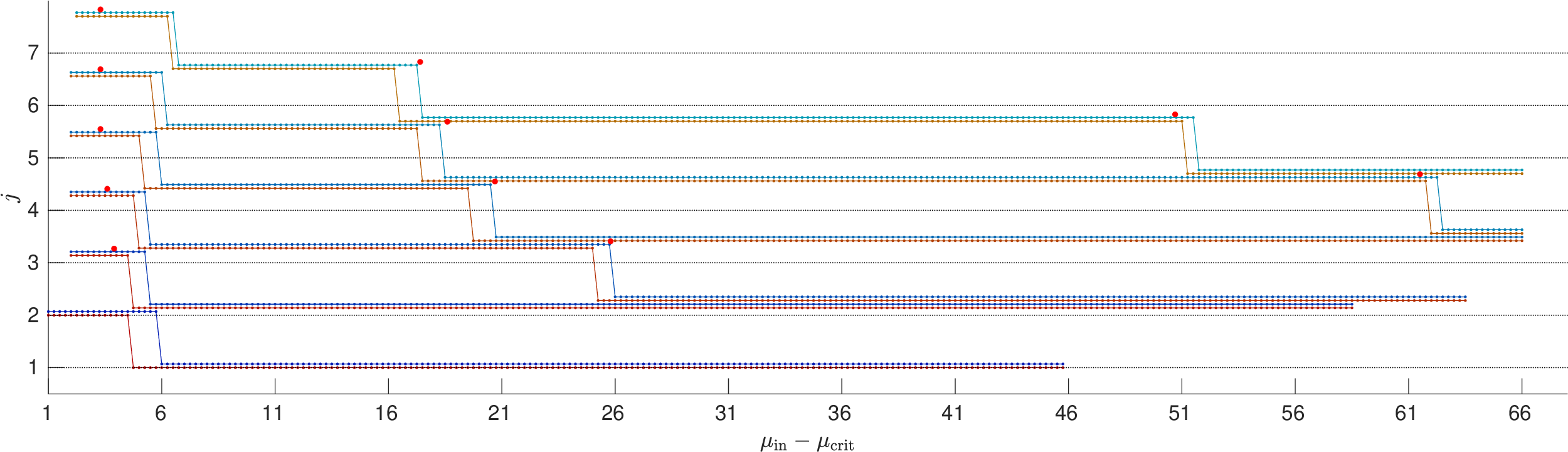}\\
    \includegraphics[width=\textwidth]{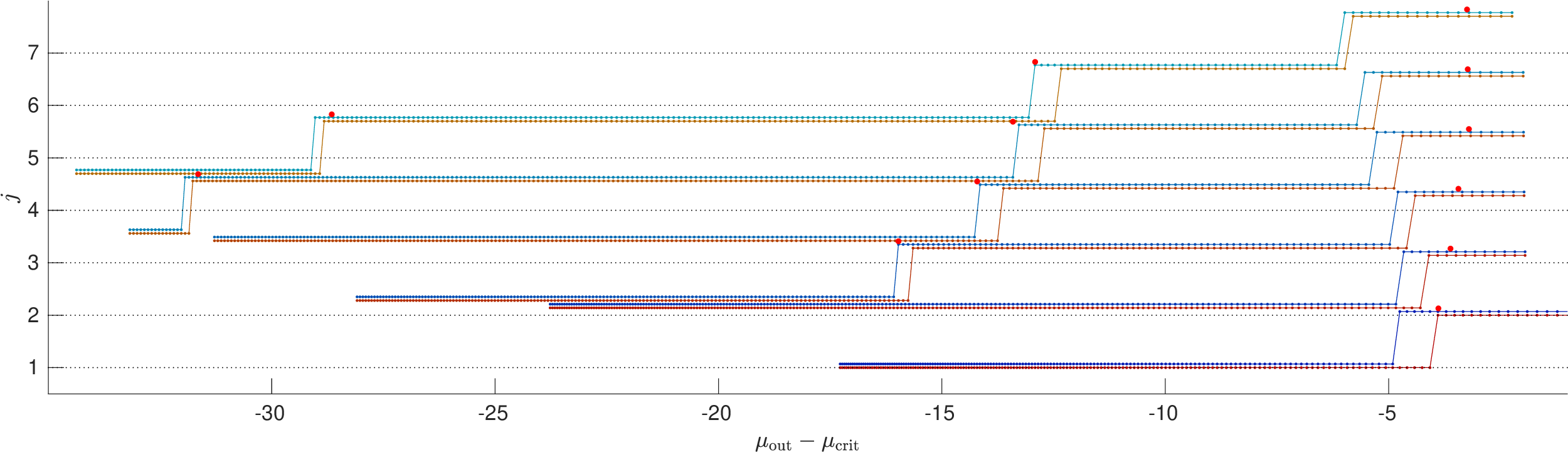}
    \caption{Drop-to numbers for initial patterns with $j=3,\ldots,8$ varying $\mu_\mathrm{in}$, plotted against $\mu_\mathrm{in}-\mu_\mathrm{c}$ (top) and  $\mu_\mathrm{out}-\mu_\mathrm{c}$ (bottom). Shown are both local (blue) and global drops (yellow), which always yield an equal drop or a drop by one less than the local drop. The actual drop-to number is the integer part of the plotted value (which is slightly shifted to improve readability). Red markers indicate the theoretical prediction for drops.  }\label{f:hd}
\end{figure}

It turns out that $b_4^{2,2}$ maximizes the amplitudes, and we can find the value of the transition from a drop-by-3 to a drop-by-4 by solving $b^4_{2,2}(\mu_\mathrm{out})=0$ together with $b^3(\mu_\mathrm{out})=0$ for $\mu_\mathrm{out}$ and $\mu_\mathrm{in}$. For finite $j$, this equation can easily be solved numerically and the predicted drop transitions compare well with numerical simulations as shown in 
Fig.~\ref{f:hd}.

Generalizing to higher transitions appears straightforward in principle but cumbersome. One obtains successively $\log$-amplitudes $b_1,b_2,b_3,b_4,\ldots,b_\ell,\ldots$ for a given $\mu_\mathrm{in}$ as functions of $\mu$, maximizing at each $\ell$ and each $\mu$ over all potential resonant source terms. Solving for $b_\ell(\mu)=0$ as a function of $\mu$ given $\mu_\mathrm{in}$ then yields the  local drop parameter value $\mu_\mathrm{out}$. The argmax of all the exit values $\mu_\mathrm{out}$ then determines the local drop number for a given $\mu_\mathrm{in}$. We caution, however, that at some point the skew product structure may not be preserved since, for instance, terms of the form $a_3\bar{a_2}$ in the equation for $a_1'$ may dominate the linear term $\lambda_1a_1$.

\subsection{Drop number transitions for large \emph{j}}

In the limit of large $j$, the criteria for transitions can be evaluated explicitly. Using the expansion for eigenvalues~\eqref{e:lamlj}, we can find the resonances as defined in~\eqref{e:rest} explicitly. Setting $\mu=3j^2-\frac{1}{2}+\hat{\mu}$, we have in decreasing order,
\[
\hat{\mu}_{1^2:2}=-3, \quad 
\hat{\mu}_{1^3:3}=-6, \quad 
\hat{\mu}_{1,2:3}=-\frac{15}{2},  \quad 
\hat{\mu}_{1,3:4}=-14, \quad
\hat{\mu}_{1,4:5}=-\frac{45}{2},
\]
and
\[
\hat{\mu}_{1^4:4}=-10, \quad 
\hat{\mu}_{1^2,2:4}=-\frac{114}{10}, \quad 
\hat{\mu}_{2^2:4}=-\frac{27}{2}, \quad 
\hat{\mu}_{1,3:4}=-14, \ldots
\]
Evaluating the log-amplitudes in~\eqref{e:b1}--\eqref{e:b4} explicitly for eigenvalues as in~\eqref{e:lamlj}, we also find explicit crossover points $\hat{\mu}_{\mathrm{in},1\to 2}$, $\hat{\mu}_{\mathrm{out},1\to 2}$,   for drop-by-1 to drop-by-2 and  $\hat{\mu}_{\mathrm{in},2\to 3}$, $\hat{\mu}_{\mathrm{out},2\to 3}$ for drop-by-2 to drop-by-3 as
\begin{equation}\label{e:mut}
\hat{\mu}_{\mathrm{in},1\to 2}=3,\quad \hat{\mu}_{\mathrm{out},1\to 2}=-3, \qquad \qquad
\hat{\mu}_{\mathrm{in},2\to 3}=15,\quad \hat{\mu}_{\mathrm{out},2\to 3}=-12.
\end{equation}
Note that the change to a drop-by-3 occurs well before the amplitude  $a_4$ becomes relevant. 

The drop values $ \hat{\mu}_{\mathrm{out},\ell}$ for a drop by $\ell$ as a function of $\hat{\mu}_\mathrm{in}$ are given by
\begin{align}
    \hat{\mu}_{\mathrm{out},1}&=-\hat{\mu}_\mathrm{in}, &\qquad 0<\hat{\mu}_\mathrm{in}<\hat{\mu}_{\mathrm{in},1\to 2},\\
     \hat{\mu}_{\mathrm{out},2}&=\frac{1}{2} \left(-3 - \sqrt{-9 + 2 \hat{\mu}_\mathrm{in}^2}\right), &\qquad \hat{\mu}_{\mathrm{in},1\to 2}<\hat{\mu}_\mathrm{in}<\hat{\mu}_{\mathrm{in},2\to 3},\\
     \hat{\mu}_{\mathrm{out},3}&=\frac{1}{6} \left(-24 - 2\sqrt{3} \sqrt{-33+ \hat{\mu}_\mathrm{in}^2}\right), &\qquad \hat{\mu}_{\mathrm{in},2\to3 }<\hat{\mu}_\mathrm{in}<\hat{\mu}_{\mathrm{in},3\to4 }.
\end{align}
For the drop-by-3 to drop-by-4 transition, we find, based on the source term $a_2^2$ and the associated 2,2:4-resonance, 
\begin{align}
\hat{\mu}_{\mathrm{in},3\to 4}&=3 \sqrt{\frac{1}{2} \left(159+28 \sqrt{14}\right)}\nonumber\\
&=34.4521\ldots,\nonumber\\
\hat{\mu}_{\mathrm{out},3\to 4}&=\frac{1}{4} \left(-30-\sqrt{2 \left(9 \left(159+28
   \sqrt{14}\right)-297\right)}\right)\nonumber\\
      &=-23.6125\ldots.\label{e:mut2}
\end{align}
The transition associated with the resonant term $a_1a_3$, is located at
\begin{align}
\hat{\mu}^{13}_{\mathrm{in},3\to 4}&=\frac{1}{2} \sqrt{21 \left(147+8 \sqrt{201}\right)}\nonumber\\
&=36.9757\ldots,\nonumber\\
\hat{\mu}^{13}_{\mathrm{out},3\to 4}&=\frac{1}{4} \left(-30-\sqrt{21 \left(147+8
   \sqrt{201}\right)-519}\right)\nonumber\\
      &=-25.0887\ldots,\nonumber
\end{align}
a larger value of $\mu_\mathrm{in}$. The drop values for finite $j$ have significant corrections, in particular for larger drop numbers or when the drop parameter value is close to the existence boundary. 

\section{Discussion}
We analyzed the slow passage through an Eckhaus bifurcation in a bounded domain. Under the conceptual assumption, well corroborated in simulations, that for perturbations of unstable patterns, the dominant Fourier mode of the perturbation determines the global drop, we reduced the analysis to the dynamics in a finite-dimensional center-manifold. We found that the drop number then depends on how far the initial parameter value is from criticality. For small distances, one observes a drop-by-1, but for larger distances the drop number increases. We derived criteria for the drop time and the drop number in general for drops up to 3, and found explicit formulas in the case of a large domain (large $j$ in our scaling). The formulas do not show an obvious pattern that would generalize to larger drop numbers, although our approach can in principle be pursued beyond drops by 3 and the transition to drops by 4. The main complicating factor in the analysis is the presence of spatio-temporal resonances, which lead to nonlinear coupling between amplitudes of modes for different drops. Spatial resonances allow for the presence of those nonlinear terms; temporal resonances determine cross-over points when growth in higher modes exceeds the resonant feeding from lower modes. 

Some avenues that would be interesting to study further are a generalization to higher drop numbers, a rigorous justification of the center manifold analysis when spectral gaps fail, and the identification of relevant higher-order terms for moderate values of $\eps$. On the other hand, it seems natural to pursue this analysis in an unbounded domain, where the Eckhaus instability is caused by essential spectrum. In this setting, the expansion of eigenvalues in the large-$j$ limit that we used throughout is universal beyond the Ginzburg-Landau equation and reflects the sideband nature with modulation equations of the form
\begin{equation}\label{e:wn}
a_t=-(a_{xx}+\hat{\mu}a - a^2)_{xx},
\end{equation}
for wavenumber corrections $a$; see for instance~\cite{vanharten}.
From this perspective, the precise form of parameter variation is irrelevant, and paths of the form $(\mu(\tau),j(\tau))$ would lead to equivalent results. The quadratic resonances, key to the calculation of drop times, are induced by the nonlinearity $(a^2)_{xx}$. It would be interesting to relate the universal quadratic term here to the quadratic coefficients modeling resonances. In this direction, it appears tempting to analyze more generally potential resonance orderings and the ensuing sequence of transitions to higher drop numbers, as outlined in \S\ref{s:hd}. Conceivably, this could be explicit in the limit as $j\to\infty$ and potentially conclude with a complete description of possible staircases as shown in Fig.~\ref{f:staircase}. On the other hand, the perspective of a wavenumber modulation equation~\eqref{e:wn} rather than an amplitude modulation such as Ginzburg-Landau demonstrates that \emph{local} drop numbers should be largely independent of the model considered and predictions, suitably adjusted by computing the relevant eigenvalues $\lambda_j(\tau)$ along a parameter path would still hold. Exploring this systematically for instance in a simple Swift-Hohenberg equation
\begin{equation}
u_t=-(k(\tau)\partial_{xx}+1)^2 u + \mu(\tau) u - u^3,
\end{equation}
in a large periodic domain, with for instance $0<\mu(\tau)\equiv \bar{\mu}\ll1$ and $k=k_0-\eps \tau$, $k_0\sim 1$, or even more realistic reaction-diffusion models such as Klausmeier's or the Gray-Scott model~\cite{klausmeier,gsk} would be a natural next step.

In this setting of an unbounded domain, rather than describing the global evolution by a heteroclinic orbit, one could start to analyze the spatio-temporal evolution of the Eckhaus instability with frozen parameter in terms of spreading fronts, and, in a second stage, incorporate the effect of spatio-temporally slowly varying parameters; see~\cite{goh2022fronts} for the effect of slowly varying environments on spatial spreading,~\cite{goh2023growing} for a review of the effect of spatio-temporal changes on pattern evolution, and~\cite{fhs_2017} for the role of spatio-temporal resonances in spatial spreading. 

In a different direction, one would also be interested in instabilities different from the Eckhaus side-band instability, in particular spatial period-doubling~\cite{siteur}, and in the effect of various types of spatio-temporal noise and variation. Even adding boundary conditions different from the simple periodic setting considered here may well destroy the simple pitchfork nature of the Eckhaus instability; see ~\cite{MR3376770} for a conceptual analysis of the effect of boundaries.

Finally, imperfections in time and space would certainly affect predictions made here. A delay of the bifurcation is still expected but shortened when small temporal noise is present~\cite{berglundgentz} and spatial noise may well change drop predictions in other ways.

\appendix
\section{Appendix -- numerical algorithms}

We present some details on numerical simulations performed to check Hypothesis~\ref{hyp: mode conservation}, corroborate the asymptotic formulas, and produce Figures~\ref{f:1:2} and~\ref{f:2:3}.

We simulated the Ginzburg-Landau equation using a spectral method with 32 Fourier modes in the case $j=6$ and 64 Fourier modes in the cases $j=8,12$. We monitored higher Fourier modes when comparing simulations of higher spatial resolution and found the chosen spatial resolution to be sufficient. Time stepping uses the second-order  exponential time differencing method ETD2RK from~\cite[(22)]{mc} with step sizes $0.005$ ($j=6$) and $0.001$ ($j=12$). The system is stiff because of large stable eigenvalues in the diffusion matrix and large eigenvalues due to the term involving $j^2$, which creates large stable eigenvalues for the linearization at equilibria even for small Fourier modes. Since exponential decay leads to very small amplitudes of relevant Fourier modes at criticality, we used high-precision arithmetic with sufficiently many digits $D\sim \max_t\{-p\log_{10}a_1\}$ so that $a_1^p$ is fully resolved when, for instance, investigating a drop number $p$. We confirmed that lower resolution causes round-off errors and alters results significantly.

We chose initial conditions as the basic pattern $A(x)\equiv \sqrt{\mu-j^2}$ for a given value $\mu=\mu_\mathrm{in}$ and a perturbation of size $\delta$ in the eigenvector associated with the first Fourier mode. We chose $\delta=0.1$ in Fig.~\ref{f:1:2} and $\delta=0.0001$ in Fig.~\ref{f:2:3}. Larger choices of $\delta$ lead to better agreement between local and global drop number transitions, while smaller choices of $\delta$ usually improve the agreement between $\mu_\mathrm{out}$ and predicted values.

We found local drop numbers by determining when the solution leaves a small neighborhood of the equilibrium, tracking whether $\|A-\frac{1}{2\pi}\int A\|_2<\delta$ and then finding the index of the Fourier mode with largest amplitude. We then integrated until the winding number  $A=\int_{x=0}^{2\pi} \partial_x(\Im \log(A(x)))\rmd x$ is nonzero, at which point we integrated for another 50 time units to let the trajectory converge to a neighborhood of a new equilibrium. The agreement between local and global drop numbers depends on the choice of $\delta$. We picked values as indicated which gave good agreement without trying to optimize.

To test the linear hypothesis, Fig.~\ref{fig:hypothesis 1 evidence}, we solved the initial-value problem for a collection of parameter values in the $(j,\mu)$-plane. We started with a perturbation of $E_j$ of size $\delta=0.1$ in the unstable $\ell$-mode and integrated until the trajectory reached another equilibrium, which in turn was determined by checking if $\sup \left(|A(x)|-\frac{1}{2\pi}\int A\right)<10^{-4}$. We used the same numerical parameters as above but performed the computations in standard double precision.

\end{document}